\documentclass[]{interact}
\usepackage[most]{tcolorbox}
\usepackage{subcaption}
\usepackage{url}
\usepackage[nokeyprefix]{refstyle}
\usepackage{varioref}
\usepackage{hyperref}
\usepackage{xr-hyper}
\usepackage{lmodern}
\usepackage{enumerate}
\usepackage{placeins}
\usepackage{rotating}
\usepackage{pdflscape}
\usepackage[labelsep=space]{caption}
\usepackage{adjustbox}
\usepackage{booktabs}
\usepackage{multirow}
\usepackage{svg}
\usepackage{graphicx}
\usepackage[natbibapa,nodoi]{apacite}
\usepackage[ruled,vlined]{algorithm2e}
\usepackage[font=small,labelfont=bf]{caption}
\usepackage{hyphenat}
\theoremstyle{plain}

\theoremstyle{definition}

\theoremstyle{remark}

\begin{document}

\articletype{ARTICLE}

\title{Topology-Preserving Line Densification for Creating Contiguous Cartograms}

\author{
\name{Nihal Z. Miaji\textsuperscript{a,b}, Adi Singhania\textsuperscript{a,b}, Matthias E. Goh\textsuperscript{b}, Callista Le\textsuperscript{b}, Atima Tharatipyakul\textsuperscript{a}, and Michael T.~Gastner\textsuperscript{a}\thanks{CONTACT Michael T.~Gastner. Email: michael.gastner@singaporetech.edu.sg}}
\affil{
\textsuperscript{a}Information and Communications Technology Cluster, Singapore Institute of Technology, 1 Punggol Coast Road, Singapore 828608 \\
\textsuperscript{b}Yale-NUS College, 16 College Avenue West, \#01-220, Singapore 138527
}
}

\maketitle

\begin{abstract}
Cartograms depict geographic regions with areas proportional to quantitative data. However, when created using density-equalizing map projections, cartograms may exhibit invalid topologies if boundary polygons are drawn using only a finite set of vertices connected by straight lines. Here we introduce a method for topology-preserving line densification that guarantees that cartogram regions remain connected and non-overlapping when using density-equalizing map projections. By combining our densification technique with a flow-based cartogram generator, we present a robust framework for strictly topology-preserving cartogram construction. Quantitative evaluations demonstrate that the proposed algorithm produces cartograms with greater accuracy and speed than alternative methods while maintaining comparable shape fidelity.
\end{abstract}


\section{Introduction}

\begin{figure}[ht!]
  \centering
  \includegraphics[width=\textwidth, keepaspectratio]{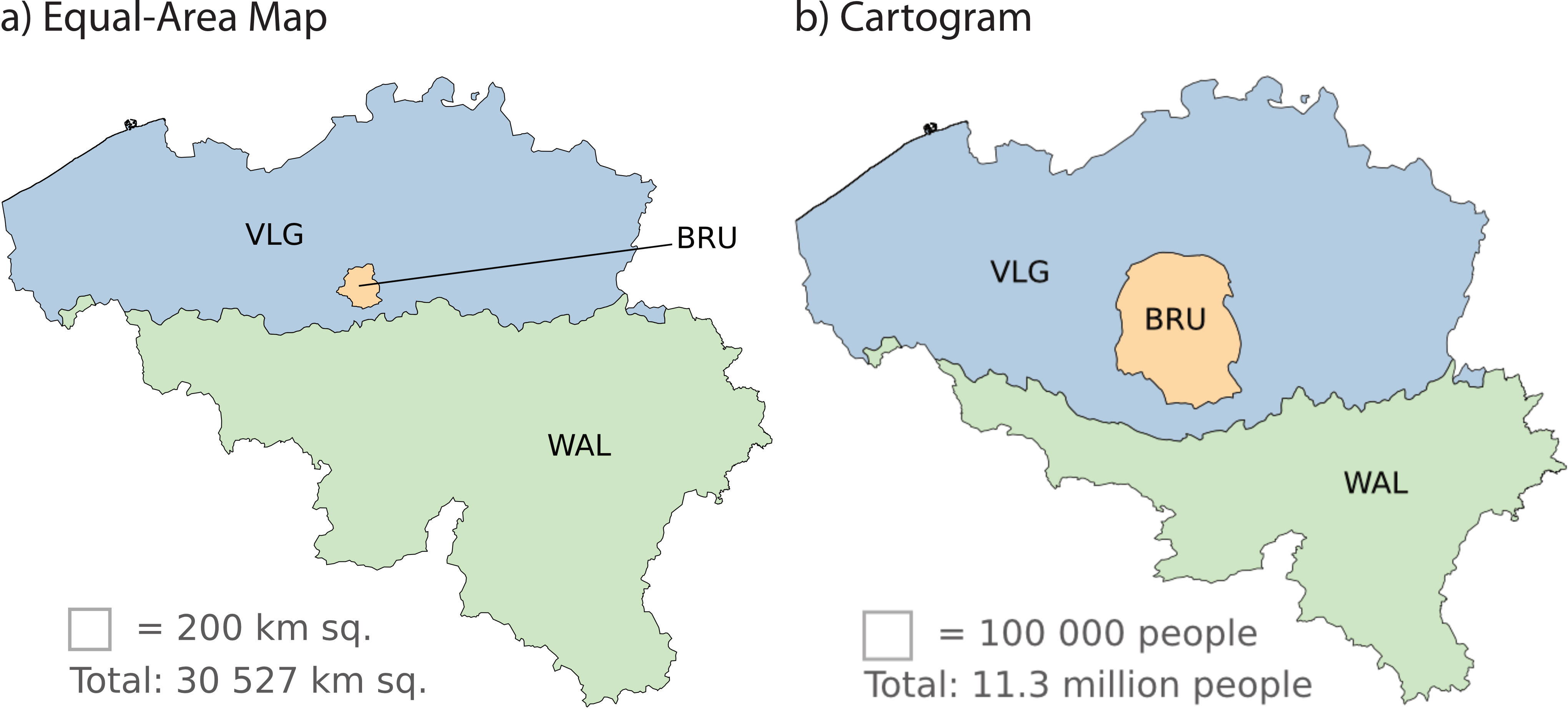}
  \caption{Maps of Belgium. a) Belgium’s regions on an equal-area map, abbreviated as BRU (Brussels), VLG (Flanders), and WAL (Wallonia). b) Contiguous cartogram representing population. As Brussels is the most densely populated region, its area is enlarged in the cartogram.}
  \label{fig:belgium_cart}
\end{figure}

Cartograms are maps in which the areas of regions (e.g., administrative divisions) are proportional to spatially extensive quantitative data (e.g., population). As cartogram areas are typically not proportional to land areas, the shapes or the contiguity of the polygons representing the geographic regions must be adjusted. If all adjacent polygons share boundaries even after resizing their areas, the cartogram is called contiguous, such as the example shown in Figure~\ref{fig:belgium_cart}.  In practice, cartograms have been applied to a wide range of data, including election results~\citep{gastner2005maps}, socio-economic indicators~\citep{dorling1996}, disease incidence~\citep{selvin1988transformations}, and climate change risks~\citep{doll2017cartograms}.

\begin{figure}[t!]
  \centering
  \includegraphics[width=\textwidth, keepaspectratio]{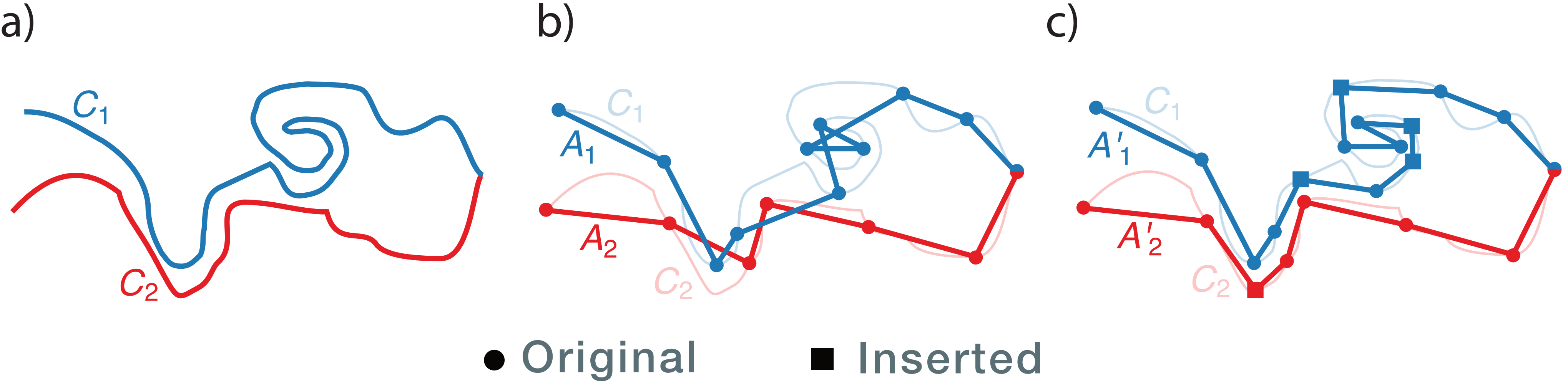}
  \caption{Illustration of line densification when approximating a curve with a finite number of points. Panel a) depicts the ground truth, showing the precise geometry of two curves, $C_1$ and $C_2$. Panel b) illustrates finite approximations, $A_1$ and $A_2$, that use a sparse distribution of points along the respective curves. Topology is violated as $A_1$ intersects itself and also $A_2$. Panel c) demonstrates line densification resulting in $A_1'$ and $A_2'$ preserving the true topology.}
  \label{fig:approximation}
\end{figure}

Contiguous cartograms can be constructed using density-equalizing map projections, as demonstrated by \citet{tobler_geographic_1963} and \citet{tobler_cartograms_2017}. However, upon transformation via a density-equalizing map projection, invalid topologies---such as intersections or gaps between adjacent polygons---can arise. This issue occurs because computer-generated maps represent enumeration units as polygons with a finite set of vertices connected by straight lines.  Here we introduce a method for topology-preserving line densification that solves this problem. Line densification involves augmenting the density of points along boundaries to allow finer representation of the underlying curves. When discretizing boundaries, line densification can ensure that neighboring regions remain connected but do not overlap, as illustrated in Figure~\ref{fig:approximation}.

To evaluate the proposed densification algorithm's effectiveness, we augment the flow-based cartogram generation algorithm by \citet{gastner2018fast} with our densification method. We refer to the resulting method as 5FCarto, where ``5F'' refers to five key properties: flow-based, faultless (i.e., no topology violations), free (open-source under AGPL-3.0), fast (quicker than other current methods), and focused (primarily allocating computational resources to regions that require significant transformation).

We compare 5FCarto to \citeauthor{sun2020applying}'s (\citeyear{sun2020applying}) force-based cartogram generation algorithm (F4Carto) and to the basic flow-based (BFB) method, which is identical to 5FCarto except that it does not include the densification step. The results, evaluated using four different performance metrics applied to 32 maps, demonstrate that 5FCarto produces topologically valid cartograms with polygon shapes comparable to those produced by other methods. Moreover, 5FCarto outperforms the alternatives in terms of area accuracy and speed, making it a robust and efficient solution for cartogram generation.

\section{Problem Statement}
\label{sec:problem_statement}

Density-equalizing map projections provide a mathematically principled method for creating contiguous cartograms. In this approach, the input data are expressed as spatial densities (e.g., population per unit area), and each region is resized so that the density becomes uniform across the entire map. Afterward, the polygon vertices are projected and connected with lines to create the cartogram. A density-equalizing map projection $\Pi$ must satisfy:

\begin{equation}
    J_\Pi(x, y) = \frac{\rho(x, y)}{\overline{\rho}}\ ,
    \label{eq:jacobian}
\end{equation}

\noindent
where $\rho$ is the density function, and $\overline{\rho}$ is the average density. The left-hand side of \Eqref{eq:jacobian} is the Jacobian determinant $J_\Pi(x, y) = \frac{\partial \Pi_x}{\partial x} \frac{\partial \Pi_y}{\partial y} - \frac{\partial \Pi_y}{\partial x} \frac{\partial \Pi_x}{\partial y}$ of the projection, which we assume to be differentiable in the entire mapping domain, for example, by applying pycnophylactic interpolation \citep{gastner2022smooth}. Additionally, we assume that the density is positive everywhere. This assumption poses no practical restriction because mixtures of positive and non-positive data (e.g., temperature anomalies or net migration) cannot be proportional to physical areas, and, thus, cartograms would be unsuitable for the visualization of such data; instead, alternative visualization techniques or appropriate data transformations must be considered. Under the assumption that $\rho(x, y)$ is strictly positive with a finite upper bound for all real-valued $x$ and $y$, Hadamard's global inverse function theorem implies that $\Pi$ is invertible~\citep{ohkita_simple_2024}. Thus, density-equalizing map projections theoretically guarantee that no intersections are generated by the transformation.

\begin{figure}[h!]
  \centering
  \includegraphics[width=\textwidth, keepaspectratio]{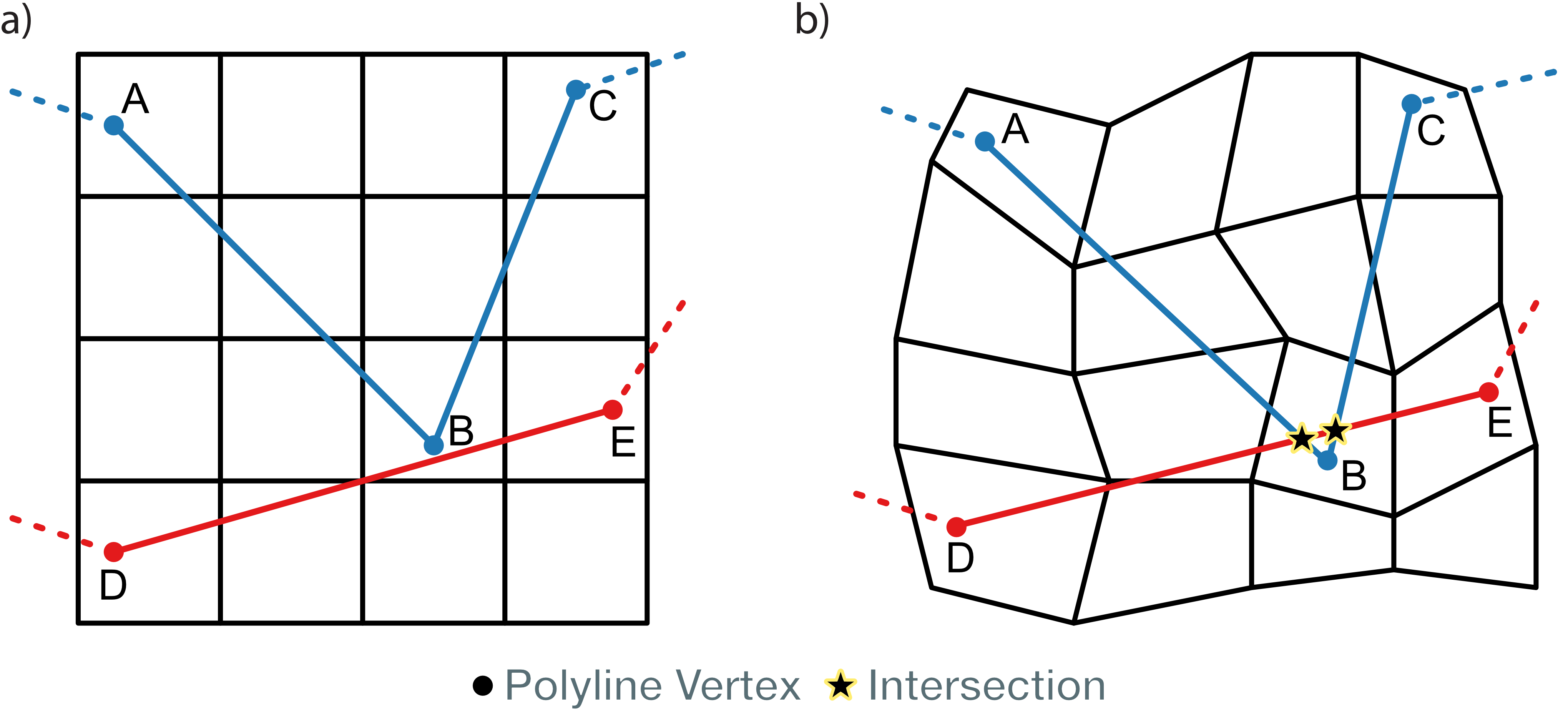}

  \caption{Illustration of a topology violation caused by projecting polylines. The figure depicts two polylines, one colored blue (ABC) and the other red (DE). Panel a) illustrates the original geometry, while panel b) shows the situation after transformation via a density-equalizing map projection. The transformed black grid lines indicate the projection. The projected positions of the polyline vertices A to E are interpolated on the basis of the quadrilaterals formed by the projected grid lines in panel b). Although the blue and red polylines do not intersect before the transformation, they cross each other afterward, resulting in an invalid topology.}
  \label{fig:intersection}
\end{figure}

In practice, however, the boundary of a region needs to be approximated by a polygon with a finite number of points. This restriction can result in topology violations emerging after the transformation, even for continuous density-equalizing map projections, as illustrated by an intersection emerging in Figure~\ref{fig:intersection}.

Our topology-preserving line densification aims to prevent such violations. To capture topology preservation precisely, we define a line approximation of a curve as a finite sequence of points on the curve (endpoints included), with each consecutive pair connected by a straight segment. For instance, \(A_1\) and \(A_2\) are line approximations of \(C_1\) and \(C_2\) in Figure~\ref{fig:approximation}. A set of line approximations \(A_1, \ldots, A_n\) for curves \(C_1, \ldots, C_n\) that have no common points, with the only possible exception of common end points, is topology-preserving if all of the following three properties hold:

\begin{enumerate}[P1:]
    \item \(A_i\) is free of self-intersections for all $i = 1, \ldots, n$.
    \item Any pair $A_i$ and $A_j$ with $i \neq j$ has no points in common except, possibly, the endpoints.
    \item At each endpoint where three or more curves meet, the approximations $A_i$, $A_j$, and $A_k$ of any three such curves preserve the same winding order as the original curves $C_i$, $C_j$, and $C_k$.
\end{enumerate}

In Figure~\ref{fig:approximation}, $A_1'$ and $A_2'$ preserve topology, whereas $A_1$ and $A_2$ do not.

\section{Related Work}

\subsection{Topology Violations in Cartograms}
\label{sec:related_work_on_topo_violations}

The problem of topology violations arising from cartogram construction is well-documented in the literature. \citet{dougenik1985algorithm} noted that their force-based algorithm can produce intersecting borders, especially around panhandle-shaped regions. Similarly, \citet{house1998continuous} observed the possibility of intersections during cartogram transformations. As a mitigation strategy, they introduced an intersection-specific penalty term in their relaxation-based method for cartogram construction. Furthermore, \citet{tobler2004thirty} acknowledged the intersection problem and suggested a technique called underrelaxation, which reduces the displacements caused by the cartogram transformation. However, Tobler noted that, while underrelaxing alleviates the problem, it does not entirely prevent it.

A well-known strategy in data visualization to address the intersection problem caused by coordinate transformations is line densification, as additional vertices along the unprojected polygon edges increase the flexibility of the projected polygons. Software implementations of line densification include the R functions \texttt{coord\_munch()} in the ggplot2 package~\citep{wickham_ggplot2_2016} and \texttt{smooth\_densify()} in the smoothr package~\citep{StrimasMackey2023}. These functions insert vertices at regular intervals. However, this method does not optimize the number of insertions. On the one hand, if too few vertices are inserted, the densification might be too sparse to prevent intersections completely. On the other hand, excessive densification causes unnecessary computational and memory overhead as well as bloated image file sizes.

\citet{dougenik1985algorithm} described a method guaranteeing non-overlapping regions when using their force-based cartogram algorithm. Their method proceeds by dividing the unprojected polygons into convex subpolygons. Yet, as pointed out by \citet{sun2020applying}, the complexity of this method has prevented its implementation in cartogram software. Instead, \cite{sun_fast_2013} suggested using quadtrees as auxiliary data structures and inserting additional vertices at intersections between polygons and quadtree edges. However, it is not self-evident how to interpolate projected coordinates such that the projection is continuous at the boundaries of adjacent quadtree cells, which may differ in size. Therefore, in this paper, we describe an algorithm that supplements the quadtree representation of the density-equalizing map projection with a Delaunay triangulation to guarantee topology preservation.

\subsection{Cartogram Evaluation}

The evaluation of cartograms lacks a universally accepted set of metrics, with different studies adopting various approaches. \citet{nusrat2016evaluating} proposed three key metrics: area error, deformation, and topological accuracy, the latter defined as the fraction of adjacent regions in the original map that are not adjacent in the cartogram. However, this metric is uninformative for contiguous cartograms because topological accuracy is always perfect. Instead, \citet{sun2020applying} extended the evaluation framework by considering topological integrity, reporting the number of overlapping or self-intersecting regions as another measure of cartogram quality. Additionally, both \citet{gastner2018fast} and \citet{sun2020applying} used area error, deformation, and running time as primary metrics for evaluating cartograms. In this paper, we follow the evaluation strategies proposed in previous works by adopting topological integrity, area error, deformation, and running time as quality metrics.

For area error, the choice of metric varies between studies. \citet{henriques2009carto} suggest traditional metrics such as the quadratic mean error, the weighted mean error, and the simple average error. \citet{sun2020applying} utilized the area-weighted mean error, which prioritizes errors from larger regions. In contrast, \citet{gastner2018fast} employed the maximum relative area error, focusing equally on all regions regardless of size. The maximum relative area error represents a stricter criterion, as the area-weighted mean error can be significantly reduced by only focusing on large regions. Furthermore, reducing the maximum relative area error not only involves reshaping large polygons but also requires modifying small polygons, a process that is computationally challenging and, thus, indicative of algorithmic robustness. Consequently, in this study we adopt the maximum relative area error as a metric.

For deformation analysis, \citet{gastner2018fast} utilized Tissot's indicatrix. However, \citet{sun2020applying} highlighted that Tissot's indicatrix is a local measure and unsuitable for measuring the global distortion of polygons. Consequently, \citet{sun2020applying} proposed using the symmetric difference---previously applied by~\citet{cano_mosaic_2015}---along with the Fréchet and Hausdorff distances as distortion metrics. In contrast, \citet{dayasagar_cartograms_2014} quantified shape errors in terms of the shapiness index, which encapsulates overall morphological similarity. Because our primary concern is the displacement of individual polygon vertices, we adopt the deformation metrics introduced by \citet{sun2020applying}. Unlike their approach of averaging the Fréchet distance, Hausdorff distance, and symmetric difference into a single composite measure, we evaluate these metrics separately to provide a more detailed and granular evaluation.

\section{Proposed Algorithm}
Algorithm~\ref{alg:proposed_densification} provides a concise summary of our densification procedure. In the following sections, we detail each step of the algorithm.

\tcbset{
    fancyalgorithmstyle/.style={
        enhanced,
        colframe=blue!50!black,
        colback=blue!5,
        coltitle=white,
        colbacktitle=blue!50!black,
        fonttitle=\bfseries,
        boxrule=1mm,
        drop shadow=black!50,
        arc=0.5mm,
        top=2mm,
        bottom=2mm
    }
}

\newcounter{algo}
\refstepcounter{algo}
\begin{tcolorbox}[fancyalgorithmstyle, title=Algorithm \thealgo: Proposed Densification]
\label{alg:proposed_densification}
\SetAlgoLined
\SetKwInput{Input}{Input}
\SetKwInput{Output}{Output}

\Input{Multipolygons representing geographic boundaries (possibly with holes), densities, and density-equalizing map projection.}
\Output{Densified multipolygons guaranteed to remain free of intersections under the density-equalizing map projection.}

\begin{enumerate}
\item \textbf{Quadtree Construction:} \\
   Recursively split leaf nodes with the largest density differences until a predetermined number of leaf nodes is reached. Then grade the quadtree to ensure that the depth difference between neighboring leaf nodes is $\leq 1$.

\item \textbf{Delaunay Triangulation:} \\
   Triangulate the quadtree corners and select the triangulation configuration that maximizes the minimum angles of the projected triangles.

\item \textbf{Line Densification:} \\
   Insert points at the intersections of polygon edges with Delaunay triangle edges.
\end{enumerate}

\end{tcolorbox}

\subsection{Quadtree Construction}\label{subsec:quadtree_construction}
\begin{figure}[ht!]
  \centering
  \includegraphics[width=\textwidth, keepaspectratio]{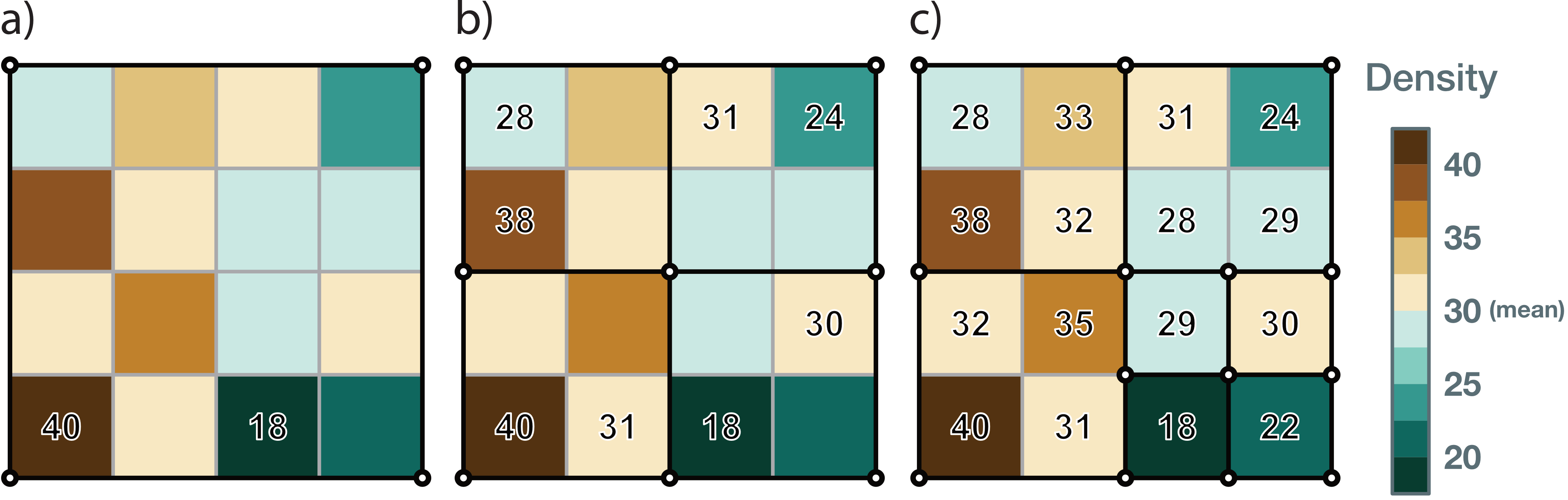}
  \caption{Illustration of the quadtree splitting criteria. The squares that are delineated by thin gray lines indicate a rasterized density grid, while those outlined in black denote quadtree leaf nodes. The quadtree is constructed by recursively splitting an $n \times n$ cell containing the largest difference between maximum and minimum density into four $(n/2) \times (n/2)$ cells until a predetermined number of leaf nodes (7 in this example) is reached. The squares in the raster grid that contain the maximum and minimum density values are annotated with those values. Panel a) depicts the initial $4 \times 4$ cell, which is split due to having the largest internal density difference (22) and because the current number of leaf nodes (1) has not reached the target (7). Panel b) illustrates the resulting subdivision with 4 leaf nodes, where again the node with the largest density difference (12, bottom-right) is selected for further splitting, as the target number of leaf nodes has still not been met. Panel c) shows the final subdivision with 7 leaf nodes, at which point the desired number of leaf nodes has been reached, marking the completion of the quadtree.}
  \label{fig:quadtree_explanation}
\end{figure}

We begin by constructing a quadtree from the rasterized input density. Following \cite{gastner2004diffusion}, any raster cell falling outside the map’s polygons is assigned the mean density, denoted by $\bar{\rho}$ in \Eqref{eq:jacobian}. We initialize the quadtree with its root as the only leaf node---a square that encloses the entire map.  We then iteratively split the leaf node containing the highest difference between maximum and minimum density until a predetermined number of leaf nodes is reached (Figure~\ref{fig:quadtree_explanation}). In our implementation, we set this target to $2^{-8} \approx 0.004$ times the number of raster cells, typically resulting in 200 to 17,000 leaf nodes, depending on the resolution of the density grid provided as input. This threshold was chosen to optimize average running times across a test set of 32 real-world maps; we report detailed measurements in the online supplement (Table \ref{tab:quadtree_depth_appen} and Figure~\ref{fig:quadtree_depth}).

After reaching the target leaf count, we perform additional splits to ensure that the depth difference between neighboring quadtree nodes does not exceed 1, a process known as grading. The role of grading for our algorithm is discussed in Section~\ref{subsec:line_densification}.

By subdividing quadtree nodes where density variations are greatest and bounding the total number of leaves, this construction concentrates computational effort on region requiring the strongest deformation while ensuring efficient execution. 

\subsection{Delaunay Triangulation}\label{subsec:delaunay_triangulation}

\begin{figure}[ht!]
  \centering
  \includegraphics[width=\textwidth, keepaspectratio]{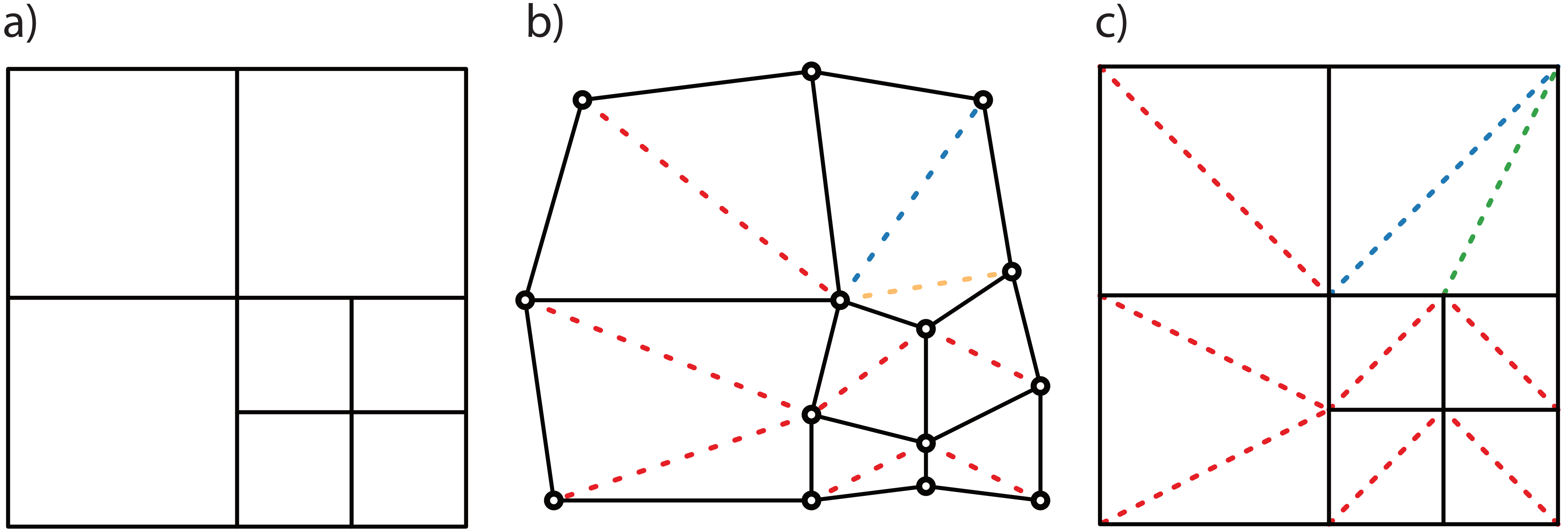}
  \caption{Illustration of the triangulation process. a) Original graded quadtree, as obtained in Figure~\ref{fig:quadtree_explanation}. b) Quadtree corners projected according to the density grid from Figure~\ref{fig:quadtree_explanation}. The projected quadtree cell edges (black solid lines) serve as constraints for computing a Delaunay triangulation, $T_{\text{temp}}$, of the projected corners, with additional triangulation edges represented by dashed lines. c) After unprojecting the edges, the final constrained Delaunay triangulation, $T$, comprises both the solid black lines and the red, green, and blue dashed lines. Although the yellow dashed line represents the Delaunay triangulation of the transformed quadtree, we do not insert it into the original quadtree because it overlaps with a boundary of a quadtree leaf node. Thus, we initially only add the blue dashed line as a constraint in the upper right leaf node before adding the green dashed line to complete the triangulation under the specified constraints.}
  \label{fig:triangulation_explanation}
\end{figure}

Once the quadtree is constructed, we generate a constrained Delaunay triangulation---denoted $T$---of the graded quadtree cell corners, as illustrated in Figure~\ref{fig:triangulation_explanation}. This triangulation is a variant of the standard Delaunay triangulation that enforces the inclusion of a predefined set of edges. Initially, we create a temporary constrained Delaunay triangulation, $T_\text{temp}$, using the projected quadtree cell corners and adding the projected quadtree cell edges as constraints.  By constructing the Delaunay triangulation in the projected space, we maximize the minimum angle in the triangle, thereby ensuring numerically robust interpolation by preventing excessively skinny triangles. Next, we unproject each edge in $T_\text{temp}$. If no other quadtree cell corner lies along an unprojected edge, we incorporate it as a constraint in $T$. After all constraints have been added, we perform a Delaunay triangulation on the unprojected quadtree vertices to construct $T$. Our implementation leverages CGAL's Delaunay triangulation module \citep{cgal:delaunay}.

\subsection{Line Densification}\label{subsec:line_densification}

\begin{figure}[ht!]
  \centering
  \includegraphics[width=\textwidth, keepaspectratio]{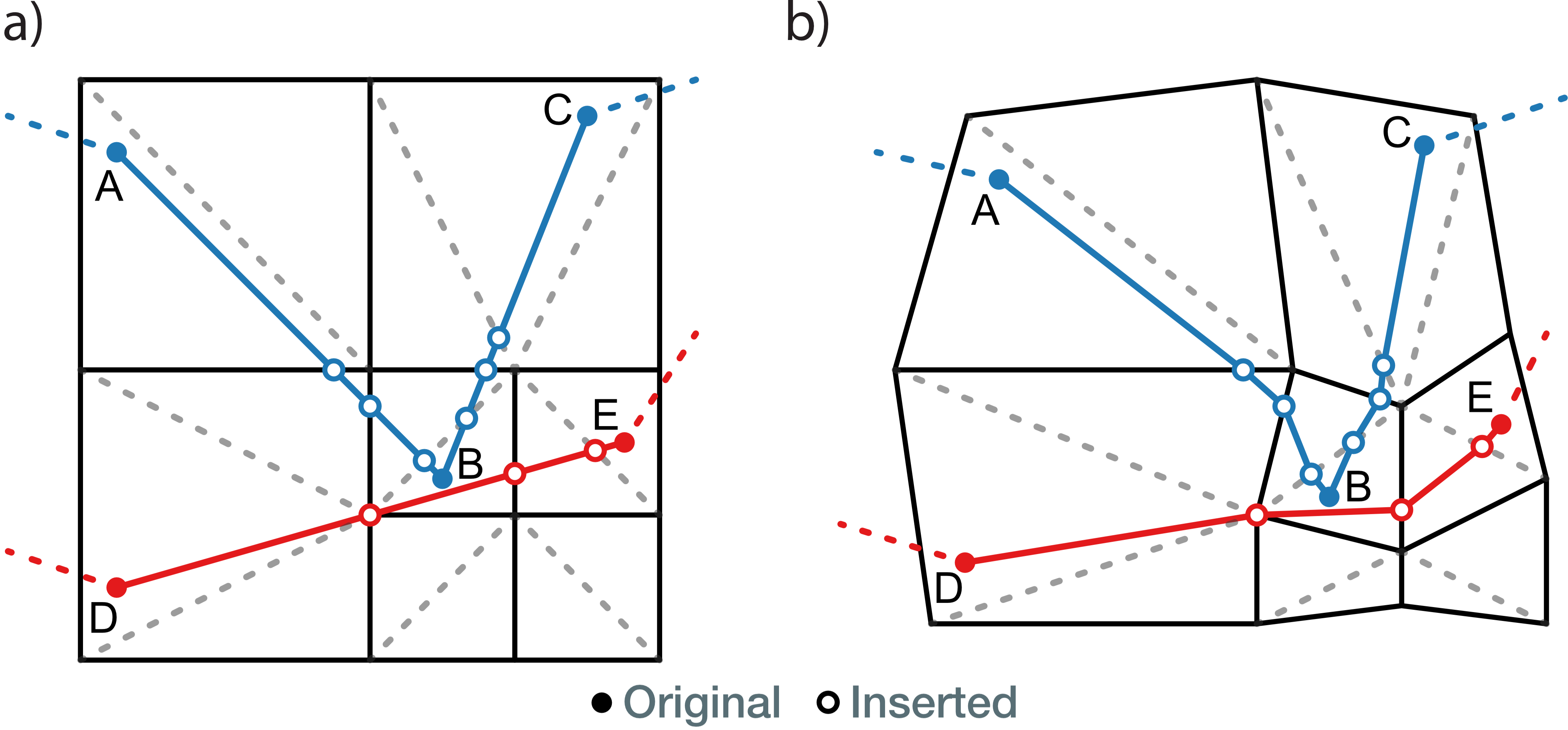}
  \caption{Illustration of our algorithm designed to prevent polyline intersections. a)~The positions of the polyline vertices are identical to those in Figure~\ref{fig:intersection}; the quadtree and the Delaunay triangulation match those in Figure~\ref{fig:triangulation_explanation}. Additional polyline vertices, represented by open circles, are added at intersections with the Delaunay triangle boundaries. b)~When the vertices added in a) are transformed using the same density-equalizing map projection as in Figure~\ref{fig:intersection}, alongside vertices A to E, the red and blue polylines do not intersect. Thus, in contrast to the earlier situation, the original topology is preserved.}
  \label{fig:densified_and_fix}
\end{figure}

After the constrained Delaunay triangulation $T$ is created, we insert points at all intersections of polygon edges with any Delaunay triangle boundary (Figure~\ref{fig:densified_and_fix}a). The projected coordinates of the original and added points are then interpolated using the uniquely defined piecewise affine transformation that maps the unprojected Delaunay triangles to their respective projections, as shown in Figure~\ref{fig:densified_and_fix}b.

\begin{figure}[ht!]
  \centering
  \includegraphics[width=\textwidth, keepaspectratio]{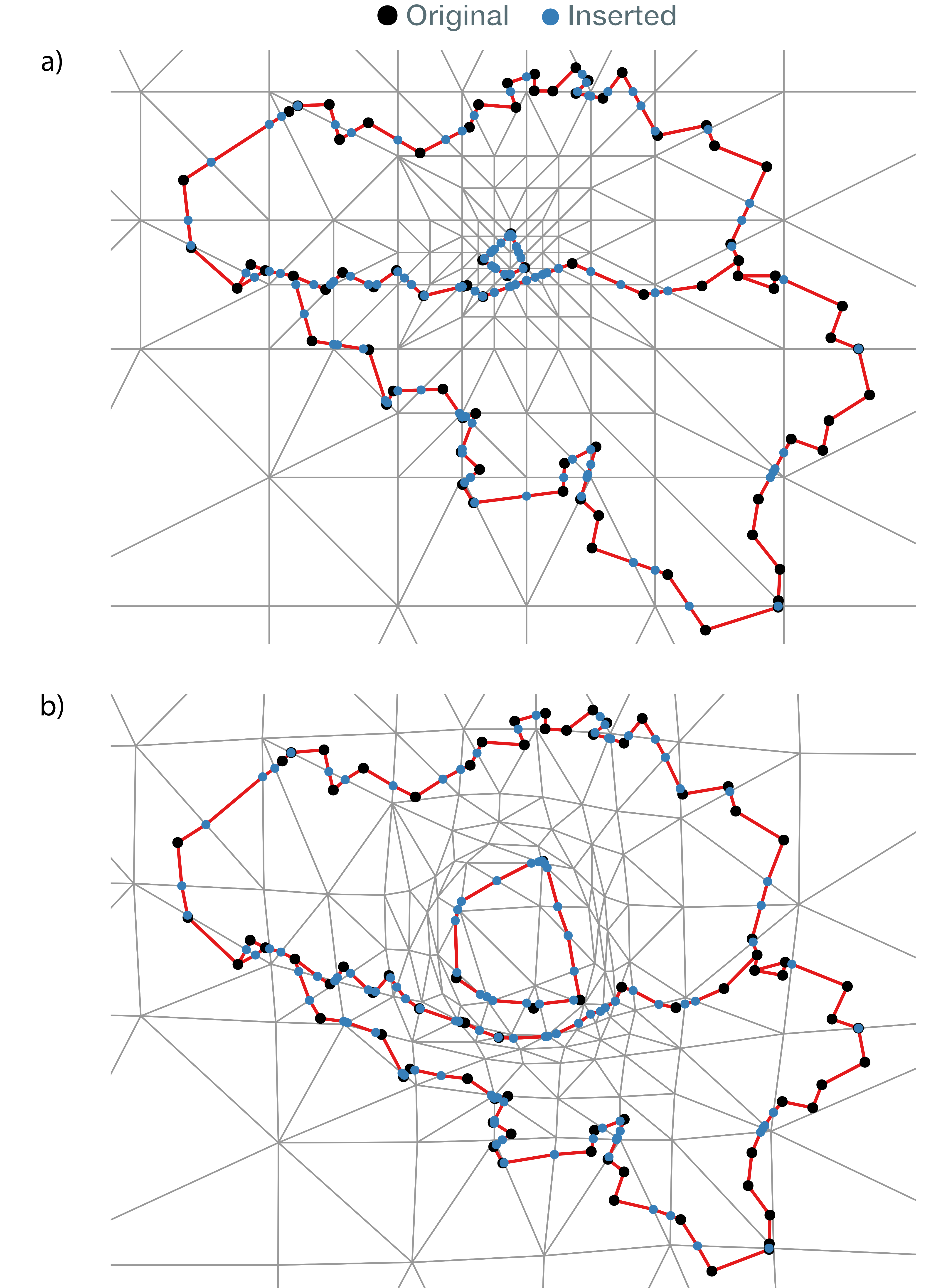}
  \caption{Illustration of Belgium's constrained Delaunay triangulation. Among the various possible triangulations of the quadtree corners in panel a), the selected triangulation maximizes the minimum angle of the projected triangles in panel b).}
  \label{fig:belgium_del}
\end{figure}

Because the quadtree is graded, the smallest angle in $T$ is $\arctan(2) - \pi/4 \approx 18.4^\circ$.
Consequently, interpolating points within these triangles is numerically more stable than in the potentially strongly elongated triangles that can arise in the Delaunay triangulation of an ungraded quadtree, where angles can approach zero.
Figure~\ref{fig:belgium_del} illustrates the application of the line-densification algorithm to real-world data for Belgium.

Finally, once all points have been transformed, we perform topology-preserving simplification to ensure that the total number of vertices remains within bounds. In our implementation, we use CGAL's simplification module \citep{cgal:simplification} and set the target number of vertices to either 10,000 or 15 times the number of polygons, whichever is greater, providing a compromise between detail and efficiency.

\subsection{Guaranteed Topology Preservation}

To establish that 5FCarto guarantees topologically valid projected polygons, we first recall that density-equalizing map projections are invertible, as explained in Section~\ref{sec:problem_statement}. Assuming that the Jacobian is sufficiently smooth to preserve orientation in the projected triangulation, the affine mappings on adjacent Delaunay triangles align continuously, preventing any intersections along their shared edges. Furthermore, line segments inside the same triangle will not intersect after the projection because the affine transformation of the triangle has a nonzero determinant, establishing properties P1 and P2. Finally, because property P3 follows directly from the orientation--preserving nature of the triangulation, the projected polygons are topologically valid.

\section{Evaluation}

\begin{figure}[ht!]
  \centering
  \includegraphics[width=\textwidth, keepaspectratio]{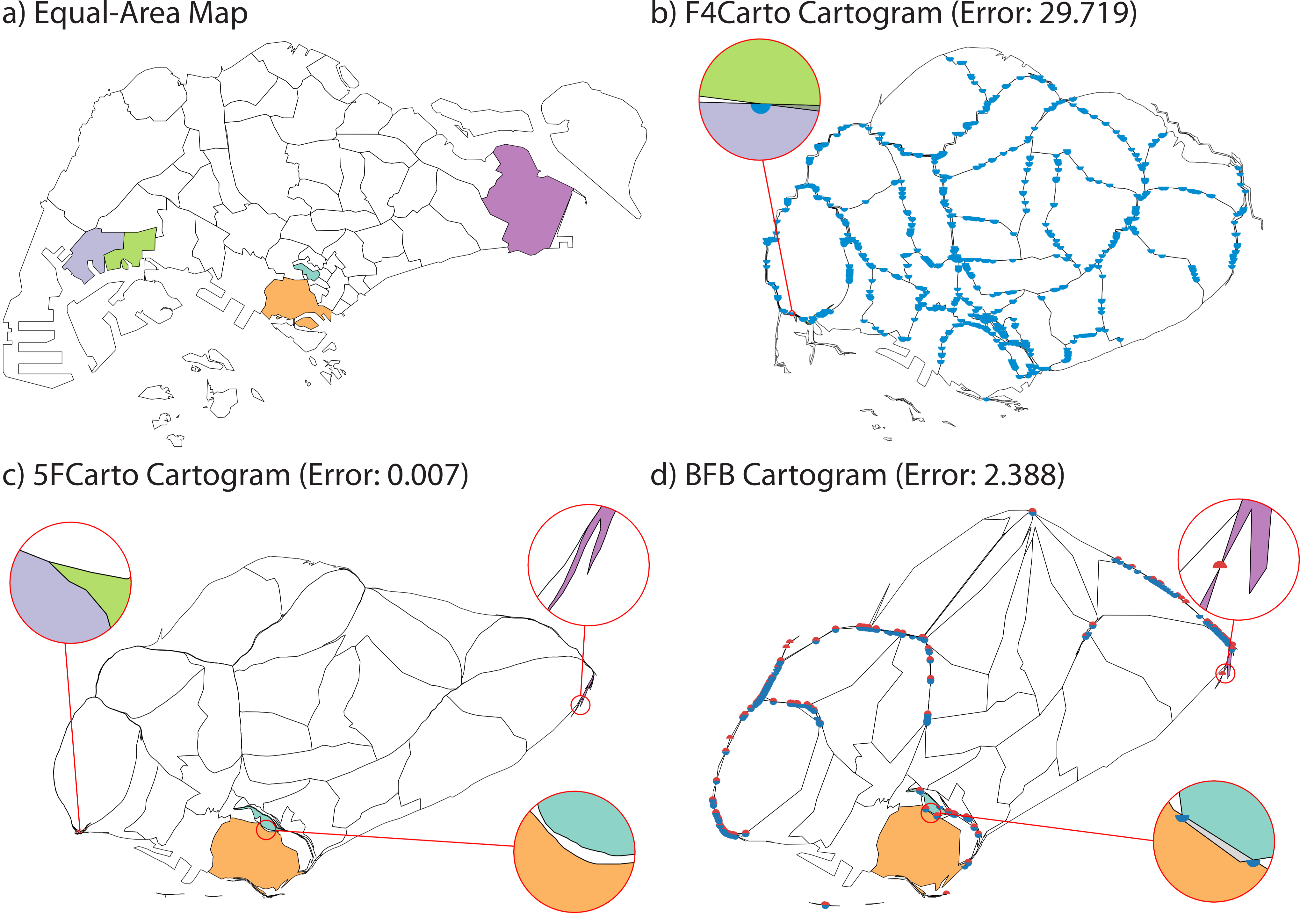}
  \caption{Illustration of Singapore's equal-area geometry, divided into planning areas, and population cartograms generated by F4Carto, 5FCarto, and BFB. The values in parentheses indicate the maximum relative area error for each cartogram. a) Equal-area map of Singapore, provided as input. b) Cartogram generated by F4Carto, with 483 overlapping intersections indicated by blue semicircles. This cartogram exhibits a high maximum relative area error, and it produces nearly rectilinear boundaries that deviate significantly from the true geographic contours. c) Cartogram generated by 5FCarto, with no intersections present. d) Cartogram generated by BFB, displaying 316 self-intersections (red upper semicircles) and 423 overlapping intersections (blue lower semicircles).}
  \label{fig:comparison}
\end{figure}

To evaluate the performance of 5FCarto, we compare it with two alternative approaches: \citeauthor{sundata}'s~(\citeyear{sundata}) F4Carto software and the original BFB method by \citet{gastner2018fast}. Figure~\ref{fig:comparison} shows an equal-area map of Singapore alongside population cartograms generated by the three methods. Cartograms are generated from a set of 32 maps using all three methods, and we assess the results using four performance metrics: area error, topological integrity, deformation, and running time. For the tests, we use real-world geometries and data, deliberately selecting a diverse range of examples representative of common use cases, to evaluate the difficulty of cartogram generation comprehensively. These test cases include regions with numerous islands and holes, identical maps with different datasets, large and small geographic areas, as well as regions with complex borders.

Let $A_i^\text{actual}$ denote the actual area of a region $i$ in the input map and $A_i^\text{target}$ the desired area in the cartogram. The density of the region is defined as
\[
\rho_i = \frac{A_i^\text{target}}{A_i^\text{actual}}\ .
\]
A density value below 1 indicates that a region is meant to shrink, while a value above 1 implies enlargement. Regions with densities significantly lower or higher than 1 require more substantial boundary adjustments, making the cartogram generation process more challenging.

To quantify the difficulty of generating a cartogram for a given dataset, we compute the density disparity, defined as the ratio of the maximum to the minimum density across all regions. A higher density disparity indicates a more challenging transformation. Based on this metric, we divide our 32 maps into four groups: Group 1 contains the easiest cases, while Group 4 comprises the most difficult ones. Table~\ref{tab:summary_density} summarizes these groups, while the supplemental Table~\ref{tab:density_appen} provides detailed information for all 32 maps.

An algorithm that performs well, particularly on the more challenging groups, can be considered more robust in handling significant deformations in region boundaries.

\begin{table}
    \caption{Summary of the 32 evaluation maps. They are grouped by density disparity, defined as the ratio of maximum to minimum density. Unless stated otherwise, target areas for the cartogram are proportional to population. Further details are available in the supplemental Table~\ref{tab:density_appen}.}
    \label{tab:summary_density}
    \centering
    \adjustbox{max width=\textwidth}{%
      \begin{tabular}{lcp{0.5\linewidth}c}
\toprule
Group   & Maps & Data & Density Disparity\\
\midrule
Group 1 & 8              & Austria, Belgium (high and low resolution), England, France (by région), Japan, Germany, Vietnam & $< 100$             \\
Group 2 & 8 & Brazil, Croatia (population and COVID-19 cases), European Union (by country), Malaysia, Switzerland (population and GDP),   Tunisia             & 100--999            \\
Group 3 & 8              & Australia, Bahamas, Egypt, France (by département), Indonesia, Singapore, USA (by state; once including Alaska and Hawaii, once without them) & 1000--9999          \\
Group 4 & 8              & Algeria, China, European Union (by region), Russia, USA (by county, without Alaska and Hawaii), World (by country), World (by region, with and without Antarctica) & $\geq 10000$           \\
\bottomrule
\end{tabular}

    }
\end{table}

\subsection{Performance Metrics}
The cartograms were evaluated using four performance metrics: area error, topological integrity, deformation, and running time. In the following sections, we provide theoretical details of each metric.

\subsubsection{Area Error}

We use the maximum relative area error to measure how well the cartogram approximates the target areas. Let \(A_i^\text{cart}\) denote the actual area of region \(i\) in the generated cartogram and \(A_i^\text{target}\) the desired target area. The relative area error for each region is defined as the absolute difference between the ratio \(A_i^\text{cart} / A_i^\text{target}\) and 1:
\[
e_i = \left| \frac{A_i^\text{cart}}{A_i^\text{target}} - 1 \right|.
\]
The maximum relative area error across all regions is then given by:
\[
E_{\text{max}} = \max_i \, e_i.
\]
A value of \(E_{\text{max}} = 0\) indicates a perfect cartogram where every region exactly matches its target area. In practice, differences below 0.01 can be considered negligible because they are unlikely to be perceptible by the human eye.

\subsubsection{Topological Integrity}

We assess the topological integrity of a cartogram by counting two types of intersections: self-intersections and overlapping intersections. Self-intersections occur when two edges of the same polygon cross, while overlapping intersections occur when edges from different polygons intersect. Ensuring that a cartogram is free of these intersections is particularly challenging, as the process of generating a cartogram involves moving points, and even slight imprecisions can lead to unwanted overlaps or self-intersections. Moreover, once a cartogram exhibits an intersection, it can be difficult to correct. Intersections not only complicate the cartogram’s structure but also confuse viewers, who may struggle to determine the true boundaries of the affected regions, a problem impacting, for instance, the online animation by \citet{riedmann2021}.

\subsubsection{Deformation}

To evaluate the deformation between the cartogram and the original map, we employ three metrics: the Fréchet distance, the Hausdorff distance, and the symmetric difference. Since the purpose of a cartogram is to adjust areas relative to its equal-area counterpart, calculating these metrics directly on the input and output regions would not yield meaningful results. Instead, we first normalize the regions by scaling them such that each region has an area equal to 1 and by centering their centroids at (0, 0). This normalization ensures that the metrics effectively quantify how well the cartogram preserves the geographic context of the original regions, enabling users to recognize and interpret them.

The Fréchet distance provides a measure of similarity between two curves by considering both the locations of the points and the order in which they appear along the curves. A common analogy is that of a person walking a dog on a leash: the person follows one curve while the dog follows another, with both allowed to move forward but not backward. The Fréchet distance is the minimum length of leash required for both to complete their respective paths~\citep{buchin_four_2017}, thus capturing the general similarity in the shape and structure of the curves.

The Hausdorff distance measures the maximum deviation between two boundaries. Formally, it is defined as the maximum of the shortest distances from any point on one boundary to the nearest point on the other boundary. This metric is particularly useful for capturing the worst-case discrepancy between the two boundaries \citep{okabe1996exact}.

The symmetric difference measures the area that is exclusive to either of the two regions. For regions \(R_1\) and \(R_2\), the symmetric difference is given by:
\[
\text{SymDiff}(R_1, R_2) = \text{Area}(R_1 \setminus R_2) + \text{Area}(R_2 \setminus R_1).
\]
This metric provides an intuitive measure of the divergence between the two shapes by quantifying the area that does not overlap.

To extend these metrics to an entire map or cartogram, which is composed of multiple regions, we compute each metric for every corresponding pair of regions between the original map and the cartogram. Each region may consist of multiple disjoint polygons with holes. When there is a one--to--one relationship between the polygon rings of the equal-area map (both outer boundaries and holes) and those of the cartogram, deformation metrics can be computed for each polygon. However, since F4Carto sometimes removes small polygons and polygons with holes, a one--to--one matching between the disjoint polygons becomes infeasible. Consequently, for the Fréchet and Hausdorff distances, we only consider the polygon with holes having the largest area for each region. We then use the outer ring from both the equal--area map and the cartogram for calculating these distances. In contrast, no such adjustment is necessary for the symmetric difference, allowing the use of all available polygons with holes.

The overall values for each deformation metric (Fréchet and Hausdorff distances, as well as the symmetric difference) are obtained by averaging the values across all regions. This aggregated approach provides a single representative value that captures the collective deformation between the original map and the cartogram. Together, these metrics enable a quantitative assessment of how well the cartogram preserves the geographic features of the original map.

\subsubsection{Running Time}

To evaluate the efficiency of cartogram generation, we consider the total running time of the algorithm as a key metric. Generating a cartogram involves iterative adjustments to the polygonal regions, a process that can become computationally intensive, especially for maps with a high density disparity or complex boundaries. Algorithms with high time complexity may struggle to scale effectively when processing large datasets. To ensure a fair assessment, it is crucial that each algorithm is run on the same machine and, if possible, under identical conditions. Additionally, if parallelization is available, it should be enabled consistently for all algorithms; otherwise, it should be disabled for all. By measuring and reporting the total running time required for cartogram generation, we provide a clear indicator of each method's computational feasibility and its suitability for practical applications.

\subsection{Results}
In this section, we present our findings for each performance metric across all three algorithms, grouped by difficulty into four levels---Group~1 (smallest density disparities; easiest) through Group~4 (largest density disparities; most challenging)---as specified in Table~\ref{tab:summary_density}. Additionally, we describe the experimental design to provide context for our evaluation.

\subsubsection{Area Error}
F4Carto's default stopping criterion is to terminate either when the error increases compared to the previous iteration or at iteration 4---whichever occurs first. However, we observed that the error at iteration 4 tends to remain high, and \citet{sundata} notes that F4Carto may sometimes overcorrect, resulting in an initial increase in area error before subsequently improving. To address this issue, we ran all three algorithms for 100 iterations and recorded the best (i.e., minimum) maximum relative area error achieved during those iterations. Note that for some maps, F4Carto terminated before completing 100 iterations due to errors caused by invalid geometry, whereas both 5FCarto and BFB consistently ran the full course without premature termination. Table~\ref{tab:summary_area_error} provides a groupwise summary, while details on all maps can be found in the supplemental Table~\ref{tab:area_error_appen}.

The results demonstrate that, across all groups, the average of the best-case maximum relative area errors, computed per map over all iterations, is consistently lower for 5FCarto than for F4Carto. While F4Carto performs well in the easiest scenarios (Groups 1 and 2), its performance degrades significantly in more challenging cases. In contrast, 5FCarto consistently reduces errors below  $10^{-7}$ regardless of map difficulty, highlighting its robustness.

\begin{table}
\caption{Average of the best-case (i.e., minimum) maximum relative area error among the first 100 iterations for each algorithm, with maps partitioned into groups based on density disparity---an indicator of cartogram generation difficulty (Group 1 being the easiest and Group 4 the most challenging). The best performance is highlighted in \textbf{bold}. Note that F4Carto occasionally terminates early due to invalid geometry errors.}
    \centering
    \adjustbox{max width=\textwidth}{
        \begin{tabular}{lccc}
\toprule
Group & \multicolumn{3}{c}{\shortstack{Averaged Best-Case Maximum Relative Area Error\\Among First 100 Iterations}} \\
\cmidrule(lr){2-4}
     & F4Carto & 5FCarto & BFB \\
\midrule

Group 1 &     0.08 & 5.426 $\times 10^{-15}$ &   \textbf{3.594 $\mathbf{\times 10^{-15}}$} \\
Group 2 &  0.06795 & 2.595 $\times 10^{-13}$ &    \textbf{3.560 $\mathbf{\times 10^{-14}}$} \\
Group 3 &   0.9591 & \textbf{7.827 $\mathbf{\times 10^{-14}}$} &     0.09684 \\
Group 4 &    9.871 & \textbf{8.967 $\mathbf{\times 10^{-9}}$} &       1.002 \\
\bottomrule
\end{tabular}

    }
    \label{tab:summary_area_error}
\end{table}

\subsubsection{Topological Integrity}\label{subsec:topo_integrity}
To evaluate this metric, we ran all three algorithms using their default parameters, with one modification to the termination criteria for F4Carto: we increased its maximum number of iterations from 4 to 10. This change was made for two reasons. First, the area error often continued to decrease beyond the fourth iteration. Second, \citeauthor{sundata}'s (\citeyear{sundata}) example parameter files suggested to use 8 to 12 iterations. In some instances, F4Carto terminated before reaching 10 iterations, as it is configured to stop when the area error increases. By default, 5FCarto and BFB stop once the area of each region in the cartogram differs from its target value by less than 1\%. After generating the cartograms, we leveraged functionality from the C++ Boost library \citep{schaling2014boost} to compute the number of intersections. See Table~\ref{tab:summary_intersections} for a groupwise summary and supplemental Table~\ref{tab:intersections_appen} for details on all maps.

The results indicate that, as the density disparity increases, F4Carto produces an increasing number of overlapping intersections, while BFB exhibits growing numbers of both self-intersections and overlapping intersections. In contrast, 5FCarto consistently achieves zero intersections across all groups.

Additionally, we compared 5FCarto's quadtree-based strategy to ggplot2's more straightforward \mbox{\texttt{coord\_munch()}} approach (see Section \ref{sec:related_work_on_topo_violations}), which inserts extra points along every polygon edge until the distance between consecutive vertices falls below a user-specified threshold. For the sample of 32 real-world datasets in Table~\ref{tab:summary_density}, the \mbox{\texttt{coord\_munch()}} strategy introduced intersections in 13 maps, despite increasing the vertex count by an average of 431\% (median: 201\%) across all maps. In contrast, 5FCarto generated all 32 cartograms free of intersections while increasing the number of vertices by only an average of 62\% (median: 20\%).

\begin{table}
    \caption{Average number of self-intersections and overlapping intersections for each algorithm, with maps partitioned into groups based on density disparity---an indicator of cartogram generation difficulty (Group 1 being the easiest and Group 4 the most challenging).  The best performance is highlighted in \textbf{bold}.}
    \label{tab:summary_intersections}

\centering
    \adjustbox{max width=\textwidth}{

\begin{tabular}{lrrr|rrr}
\toprule
\multirow{2}{*}{Group} & \multicolumn{3}{c}{Self-intersections} & \multicolumn{3}{c}{Overlap Intersections} \\
\cmidrule(lr){2-4}\cmidrule(lr){5-7}
                   & F4Carto & 5FCarto & BFB & F4Carto & 5FCarto & BFB \\
\midrule

Group 1 &                        \textbf{0.0} &                        \textbf{0.0} &                            0.5 &                           37.9 &                           \textbf{0.0} &                               0.9 \\
Group 2 &                        \textbf{0.0} &                        \textbf{0.0} &                            0.6 &                           81.8 &                           \textbf{0.0} &                              \textbf{0.0} \\
Group 3 &                        \textbf{0.0} &                        \textbf{0.0} &                           55.8 &                          202.6 &                           \textbf{0.0} &                              61.9 \\
Group 4 &                        \textbf{0.0} &                        \textbf{0.0} &                          198.5 &                          617.4 &                           \textbf{0.0} &                             134.2 \\
\bottomrule
\end{tabular}

}
\end{table}

\subsubsection{Deformation}
For this metric, we ran all three algorithms using the stopping criteria as described in Section~\ref{subsec:topo_integrity}.
To compute the distance metrics,
we employed the C++ Boost library \citep{schaling2014boost}. Table~\ref{tab:summary_similarity} presents a groupwise summary, while supplemental Table~\ref{tab:similarity_appen} provides details on all maps.

The results indicate that the deformations produced by 5FCarto and BFB are comparable across all three deformation metrics. With the exception of the Fréchet distance, on average, F4Carto appears to perform better than both 5FCarto and BFB when density disparity is high.

However, it is important to note that deformation and area error are conflicting minimization objectives. For instance, an algorithm that functions essentially as an identity transformation would produce cartograms with zero deformation but a large area error, rendering the results practically useless. Although F4Carto is not an identity transformation, its area error upon termination remains orders of magnitude higher (5.07 for Group 3 and 39.75 for Group 4) than that of 5FCarto, which, for the purpose of this measurement, was stopped only once the area error dropped below 0.01. This observation suggests that F4Carto's cartograms undergo minimal boundary modifications and may not accurately meet the target areas, whereas 5FCarto's lower area error is achieved through more substantial boundary adjustments that may compromise the preservation of the original geographic features.

\begin{table}
    \caption{Average Fréchet distance, Hausdorff distance, and symmetric difference for each algorithm, with maps partitioned into groups based on density disparity---an indicator of cartogram generation difficulty (Group 1 being the easiest and Group 4 the most challenging). The best performance is highlighted in \textbf{bold}.}
    \label{tab:summary_similarity}
    \centering
\adjustbox{max width=\textwidth}{

\begin{tabular}{lccc|ccc|ccc}
\toprule
\multirow{2}{*}{Group} & \multicolumn{3}{c}{Average Fr\'echet Distance} & \multicolumn{3}{c}{Average Hausdorff Distance} & \multicolumn{3}{c}{Average Symmetric Difference} \\
\cmidrule(lr){2-10}
                   & F4Carto & 5FCarto & BFB & F4Carto & 5FCarto & BFB & F4Carto & 5FCarto & BFB \\
\midrule

Group 1 &                      0.66 &                      0.36 &                         \textbf{0.34} &                        0.28 &                        0.24 &                           \textbf{0.23} &                          0.61 &                          \textbf{0.47} &                             \textbf{0.47} \\
Group 2 &                      0.66 &                      \textbf{0.51} &                         \textbf{0.51} &                        \textbf{0.31} &                        0.32 &                           0.32 &                          0.68 &                          \textbf{0.64} &                             0.65 \\
Group 3 &                      \textbf{1.01} &                      1.14 &                         1.85 &                        \textbf{0.37} &                        0.42 &                           1.01 &                          \textbf{0.70} &                          0.75 &                             0.76 \\
Group 4 &                      1.16 &                      \textbf{0.95} &                         1.04 &                        \textbf{0.38} &                        0.46 &                           0.46 &                          \textbf{0.66} &                          0.83 &                             0.85 \\
\bottomrule
\end{tabular}

}
\end{table}

\subsubsection{Running Time}
Comparing running times between algorithms is challenging when their stopping criteria differ. By default, both 5FCarto and BFB continue iterating until the area error falls below 0.01, which typically results in longer running times compared to F4Carto. To ensure a fair comparison, we allowed F4Carto to terminate using the stopping criterion described in Section~\ref{subsec:topo_integrity}---that is, when its area error first increases relative to the previous iteration or at iteration 10. Then, for both 5FCarto and BFB, we continued execution until their area errors dropped below the level reached by F4Carto or, in cases where they never dropped below this level, until 100 iterations were completed.

All three algorithms were executed on the same machine to maintain a consistent testing environment. F4Carto was run natively on Microsoft Windows 11 Home using a 12th Gen Intel(R) Core(TM) i7-12700H with 16 GB of RAM. Since 5FCarto and BFB cannot run natively on Windows, they were executed within an Ubuntu container using Docker in Windows Subsystem for Linux on the same hardware. Due to the virtualization overhead, the running times for 5FCarto and BFB are expected to be slower than they would be on a native Linux system with identical hardware. Moreover, although all three algorithms support parallel processing, enabling parallelization in F4Carto often caused it to stop responding during cartogram generation. Therefore, parallelization was disabled for all algorithms to ensure a fair and stable running time comparison.
Table~\ref{tab:summary_time} provides a groupwise summary, and the supplemental Table~\ref{tab:running_time_appen} contains details on all maps.

The results indicate that, at high density disparities (Group~4), 5FCarto achieves the shortest running times of the three methods.  This observation suggests that the overhead of densification is more than offset by faster convergence to accurate areas.

\begin{table}
    \caption{Average running time and maximum relative area error at the time when F4Carto terminates, with maps partitioned into groups based on density disparity---an indicator of cartogram generation difficulty (Group 1 being the easiest and Group~4 the most challenging). Note that both 5FCarto and BFB can, in principle, achieve substantially smaller area errors, as indicated in Table~\ref{tab:area_error_appen} in the supplemental material. However, BFB failed to achieve a lower area error than F4Carto for one map in Group~4 and ran the full 100 iterations, resulting in an unusually large average time. The best performance is highlighted in \textbf{bold}.}
    \label{tab:summary_time}
    \centering
\adjustbox{max width=\textwidth}{

\begin{tabular}{lccc|ccc}
\toprule
\multirow{2}{*}{Group} & \multicolumn{3}{c}{Average Time (s)} & \multicolumn{3}{c}{\shortstack{Average Maximum Relative Area \\ Error at Termination}} \\
\cmidrule(lr){2-4}\cmidrule(lr){5-7}
                     & F4Carto & 5FCarto & BFB & F4Carto & 5FCarto & BFB \\
\midrule

Group 1 &         0.296 &         0.087 &            \textbf{0.068} &                                    0.385 &                                    \textbf{0.242} &                                       0.244 \\
Group 2 &         0.617 &         0.152 &            \textbf{0.082} &                                    0.369 &                                    0.266 &                                       \textbf{0.260} \\
Group 3 &         1.219 &         0.210 &            \textbf{0.179} &                                    5.072 &                                    \textbf{3.346} &                                       3.740 \\
Group 4 &         3.737 &         \textbf{0.621} &           30.253 &                                   39.747 &                                   \textbf{35.662} &                                      37.129 \\
\bottomrule
\end{tabular}

}
\end{table}

\section{Application: 2024 Presidential Election Cartogram of the United States}

\begin{figure}[t!]
  \centering
  \includegraphics[width=\textwidth, keepaspectratio]{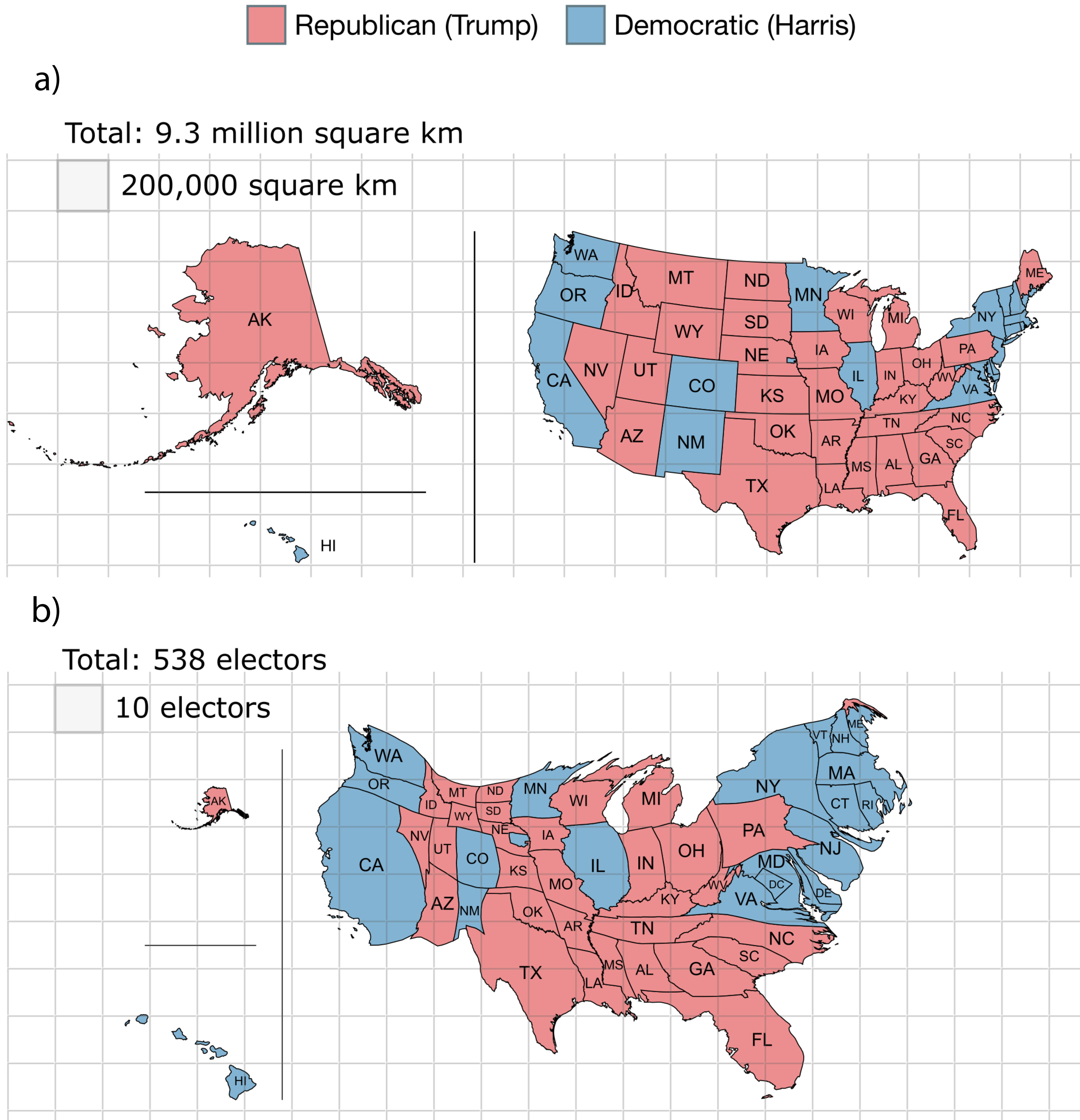}
  \caption{
    Maps of the United States. Colors represent the party that won the electoral votes in the 2024 presidential election---by state or, in Maine (ME) and Nebraska (NE), congressional district. a)~Equal-area map. b)~Cartogram generated by 5FCarto, with each region’s area proportional to its electoral-vote count.
  }
  \label{fig:us_elec}
\end{figure}

As a real-world case study, we apply the 5FCarto algorithm proposed in this paper to data from the 2024 United States presidential election. These election results have formed the basis for cartograms published by various news outlets---for example, in articles by \citet{ABCNews2024}, \citet{wsj2024}, and \citet{LeMonde2024}---because such maps eliminate visual overemphasis of land area relative to electoral weight. Figure~\ref{fig:us_elec} shows the equal-area reference map used as input to 5FCarto in panel~a) and the resulting cartogram in panel~b), where areas of all states are proportional to their number of electoral votes.

Because most states use winner-take-all rules, each state is shaded uniformly: red if the Republican ticket won the statewide popular vote, blue if the Democratic ticket won it. Maine and Nebraska are exceptions: they allocate one electoral vote to the winner of each congressional district and two at-large votes to the statewide winner. Consequently, Figure~\ref{fig:us_elec} divides both states into two regions, aggregating the congressional districts won by each party; each region is then scaled in proportion to the number of electoral votes that party is awarded in that state.

It is standard cartographic practice to display Alaska and Hawaii---both noncontiguous with the rest of the country---as insets, allowing them to be shifted closer to the other 48 states. We adopted this convention for 5FCarto, depicting both Alaska and Hawaii as insets scaled in proportion to the conterminous United States. Because this scaling is merely an affine transformation, the computational difficulty of cartogram construction is determined almost entirely by the density disparity within the conterminous United States. The resulting value (1,662.45) places this example in Group~3 of Table~\ref{tab:summary_density}, indicating a moderate computational challenge. Because F4Carto does not support insets, we excluded Alaska and Hawaii from its input and report statistics only for the conterminous United States across all algorithms.

As with all previously tested datasets, 5FCarto's output was free of polygon intersections. While F4Carto and BFB produced no self-intersections, they generated 168 and 2 overlapping intersections, respectively. Adopting the algorithms' default settings, 5FCarto was slightly faster than F4Carto (0.923 versus 0.926 seconds) but slower than BFB (0.829 seconds) due to the required densification and line simplification. Upon termination, 5FCarto reduced the maximum relative area error to $6.21\times 10^{-3}$---the same order of magnitude as BFB ($6.43\times10^{-3}$) but substantially better than F4Carto (0.744). All three algorithms yielded comparable polygon-similarity scores, measured by the Fr\'echet distance (5FCarto: 0.380; F4Carto: 0.391; BFB: 0.366), Hausdorff distance (0.276, 0.271, 0.264), and symmetric difference (0.512, 0.539, 0.517).

This example corroborates our earlier finding that, among the three algorithms tested, only 5FCarto perfectly preserves the topology. Consistent with previous results, 5FCarto accomplishes this task without substantially increasing running time, area error, or deformation.

\section{Conclusion}

In this paper, we addressed the issue of topology violations during the transformation of polylines in the construction of contiguous cartograms using density-equalizing map projections. Despite theoretical guarantees, practical implementations often result in intersections due to the approximation of region boundaries with a finite set of points connected by straight lines. We reviewed existing solutions, identified their limitations, and introduced a topology-preserving line-densification algorithm, which leverages the Delaunay triangulation of a graded quadtree to ensure that polylines transform without intersecting.

In our comparative analysis, our line-densification algorithm, 5FCarto, consistently generated cartograms with perfect topological integrity. While F4Carto and BFB may yield acceptable results for simpler maps, their performance degrades under more challenging conditions, with increasing area errors and topological violations. In contrast, 5FCarto can achieve low area errors (below $10^{-7}$ in all investigated cases; see supplemental Table~\ref{tab:area_error_appen}) while still exhibiting running times comparable to all tested alternatives (see supplemental Table~\ref{tab:running_time_appen}).

Although our algorithm performs well on challenging datasets, ultimately, as a computer algorithm, it relies on floating-point arithmetic to solve the underlying equations. Because floating-point representations have limited precision, convoluted boundaries or maps with extreme density disparities may still produce overlaps or invalid topologies due to numerical errors. Existing methods may be adapted to correct these floating-point artifacts, such as C++ CGAL’s polygon repair module \citep{cgal:polygon-repair}. Adopting higher-precision numeric types may further reduce such errors, albeit at the cost of increased computational expense.

Future research should explore opportunities for further optimization and allowing spherical input geometries alongside Cartesian ones~\citep{li2018diffusion-based, lyu2024spherical}. Nonetheless, even in its current implementation, 5FCarto's line densification represents a significant contribution to robust and efficient cartogram generation and may also be applicable to other continuous deformation visualizations in computer-generated imagery.

\section*{Acknowledgements}

We would like to thank Philemon Heng for initial code development and Edward Liau for curating the test data. We acknowledge the use of ChatGPT for language editing in the final stages of manuscript preparation. We are grateful to the anonymous reviewers whose comments improved this paper.

\section*{Funding}

This project is supported by the Ministry of Education, Singapore, under its Academic Research Fund Tier 2 (EP2) programme (Award No. MOE-T2EP20221-0007). Any opinions, findings, and conclusions or recommendations expressed in this material are those of the authors and do not reflect the views of the Ministry.

\section*{Disclosure statement}

The authors report that there are no competing interests to declare.

\section*{Data availability statement}

The data that support the findings of this study are openly available in figshare at \url{https://doi.org/10.6084/m9.figshare.28478408}.

\bibliographystyle{apacite}
\bibliography{references}

\clearpage
\clearpage

\setcounter{table}{0}
\setcounter{figure}{0}
\renewcommand{\thetable}{S\arabic{table}}
\renewcommand{\thefigure}{S\arabic{figure}}

\section*{\Large{%
 Supplementary Information}}

\begin{table}[htbp]
    \centering
    \caption{Group Details}
    \label{tab:density_appen}
    \adjustbox{max width=\textwidth}{
    \begin{tabular}{lcccc}
\toprule
Map & Group & Min Density & Max Density & Density Disparity (Ratio) \\
\midrule

France Population by Metropolitan Region (2008) & \multirow{8}{*}{Group 1} & 0.509 & 8.44 & 16.6 \\
England Population by Region (2022) &  & 0.552 & 12.87 & 23.3 \\
Belgium Population by Region (2022) &  & 0.574 & 19.91 & 34.7 \\
Simplified Belgium Population by Region (2022) &  & 0.577 & 22.88 & 39.6 \\
Germany Population by State (2011) &  & 0.313 & 15.98 & 51.0 \\
Austria Population by State (2020) &  & 0.553 & 43.54 & 78.8 \\
Vietnam Population by Province (2019) &  & 0.175 & 14.66 & 83.8 \\
Japan Population by Prefecture (2020) &  & 0.197 & 19.28 & 98.0 \\
\midrule
Croatia Population by County (2021) & \multirow{8}{*}{Group 2} & 0.116 & 17.47 & 150.2 \\
Switzerland Population by Canton (2016) &  & 0.135 & 25.58 & 190.0 \\
Croatia Covid Cases by County (2022) &  & 0.099 & 19.66 & 198.7 \\
Brazil Population by State (2021) &  & 0.109 & 28.19 & 258.7 \\
Malaysia Population by State (2020) &  & 0.201 & 83.52 & 414.7 \\
European Union Population by Country (2021) &  & 0.032 & 15.29 & 472.8 \\
Switzerland GDP by Canton (2019) &  & 0.106 & 59.63 & 564.3 \\
Tunisia Population by Governorate (2014) &  & 0.055 & 51.93 & 946.4 \\
\midrule
Australia Population by State and Territory (2021) & \multirow{8}{*}{Group 3} & 0.055 & 57.08 & 1035.5 \\
Metropolitan France Population by Département (2022) &  & 0.125 & 194.37 & 1557.1 \\
Conterminous USA Population by State (2020) &  & 0.054 & 91.95 & 1700.9 \\
Bahamas Population by District (2010) &  & 0.019 & 38.47 & 1983.3 \\
Indonesia Population by Province (2024) &  & 0.031 & 109.62 & 3565.0 \\
Singapore Population by Planning Area (2015) &  & 0.001 & 5.75 & 4933.0 \\
USA Population by State (2020) &  & 0.014 & 109.25 & 8000.6 \\
Egypt Population by Governorate (2017) &  & 0.006 & 50.43 & 8697.6 \\
\midrule
European Union Population by Region (2020) & \multirow{8}{*}{Group 4} & 0.010 & 150.91 & 15301.2 \\
USA Population by County (2021) &  & 0.024 & 479.71 & 20274.7 \\
Algeria Population by Wilaya (2022) &  & 0.012 & 257.31 & 20979.3 \\
China Population by Province (2020) &  & 0.021 & 1621.51 & 77171.4 \\
World Population by Country and Region without Antarctica &  & 0.006 & 869.08 & 137532.1 \\
Russia Population by Federal Subject (2010) &  & 0.008 & 1448.55 & 173736.4 \\
World Population by Country (2010) &  & 0.006 & 2836.26 & 444839.3 \\
World Population by Country and Region &  & 0.001 & 948.66 & 799275.4 \\
\bottomrule
\end{tabular}

    }
\end{table}

\begin{table}[htbp]
    \centering
    \caption{Area Error Details}
    \label{tab:area_error_appen}
    \adjustbox{max width=\textwidth}{
    \begin{tabular}{lccc}
\toprule
Map & \multicolumn{3}{c}{\shortstack{Best-Case Maximum Relative Area Error\\Among First 100 Iterations}} \\
\cmidrule(lr){2-4}
     & F4Carto & 5FCarto & BFB \\
\midrule

                      Algeria Population by Wilaya (2022) &   0.2248 & 3.908 $\times 10^{-14}$ &    2.220 $\times 10^{-15}$ \\
       Australia Population by State and Territory (2021) &   0.5549 & 7.661 $\times 10^{-15}$ &   1.998 $\times 10^{-15}$ \\
                       Austria Population by State (2020) &   0.0853 & 3.664 $\times 10^{-15}$ &   8.882 $\times 10^{-16}$ \\
                    Bahamas Population by District (2010) &    1.012 &  2.840 $\times 10^{-13}$ &    2.920 $\times 10^{-14}$ \\
                      Belgium Population by Region (2022) & 0.007039 & 4.441 $\times 10^{-16}$ &    2.220 $\times 10^{-16}$ \\
                        Brazil Population by State (2021) &   0.2113 & 4.219 $\times 10^{-15}$ &   2.998 $\times 10^{-15}$ \\
                      China Population by Province (2020) &   0.7843 & 2.558 $\times 10^{-13}$ &   2.442 $\times 10^{-15}$ \\
              Conterminous USA Population by State (2020) &  0.02991 & 1.366 $\times 10^{-14}$ &   6.661 $\times 10^{-15}$ \\
                     Croatia Covid Cases by County (2022) &  0.01566 & 1.599 $\times 10^{-14}$ &   3.553 $\times 10^{-15}$ \\
                      Croatia Population by County (2021) &  0.01161 & 7.438 $\times 10^{-15}$ &   2.442 $\times 10^{-15}$ \\
                   Egypt Population by Governorate (2017) &   0.1998 & 2.465 $\times 10^{-14}$ &   1.998 $\times 10^{-15}$ \\
                      England Population by Region (2022) &   0.1488 & 8.882 $\times 10^{-15}$ &   3.553 $\times 10^{-15}$ \\
              European Union Population by Country (2021) &    0.187 & 6.786 $\times 10^{-13}$ &   2.519 $\times 10^{-13}$ \\
               European Union Population by Region (2020) &    1.689 & 3.484 $\times 10^{-11}$ &   9.919 $\times 10^{-11}$ \\
          France Population by Metropolitan Region (2008) &  0.01049 & 6.661 $\times 10^{-16}$ &   1.332 $\times 10^{-15}$ \\
                       Germany Population by State (2011) &     0.15 &  1.110 $\times 10^{-15}$ &   1.221 $\times 10^{-15}$ \\
                  Indonesia Population by Province (2024) &   0.1644 &  7.550 $\times 10^{-15}$ &   9.992 $\times 10^{-15}$ \\
                    Japan Population by Prefecture (2020) &   0.1077 & 2.287 $\times 10^{-14}$ &   1.399 $\times 10^{-14}$ \\
                      Malaysia Population by State (2020) &  0.05455 & 3.442 $\times 10^{-15}$ &   3.553 $\times 10^{-15}$ \\
     Metropolitan France Population by Département (2022) &   0.2119 & 4.552 $\times 10^{-15}$ &   5.218 $\times 10^{-15}$ \\
              Russia Population by Federal Subject (2010) &    1.099 & 3.264 $\times 10^{-14}$ &   3.375 $\times 10^{-14}$ \\
           Simplified Belgium Population by Region (2022) & 0.007851 &  2.220 $\times 10^{-16}$ &   8.882 $\times 10^{-16}$ \\
             Singapore Population by Planning Area (2015) &    5.284 & 2.491 $\times 10^{-13}$ &      0.7747 \\
                         Switzerland GDP by Canton (2019) &  0.02781 & 1.284 $\times 10^{-12}$ &   1.532 $\times 10^{-14}$ \\
                  Switzerland Population by Canton (2016) &  0.02186 &  7.860 $\times 10^{-14}$ &   4.108 $\times 10^{-15}$ \\
                 Tunisia Population by Governorate (2014) &  0.01385 & 3.109 $\times 10^{-15}$ &   8.882 $\times 10^{-16}$ \\
                          USA Population by County (2021) &    4.046 & 7.136 $\times 10^{-08}$ &       8.014 \\
                           USA Population by State (2020) &   0.2157 & 3.497 $\times 10^{-14}$ &   1.465 $\times 10^{-14}$ \\
                    Vietnam Population by Province (2019) &   0.1228 & 5.551 $\times 10^{-15}$ &   6.661 $\times 10^{-15}$ \\
                       World Population by Country (2010) &    7.256 &  7.630 $\times 10^{-11}$ &   5.307 $\times 10^{-14}$ \\
                   World Population by Country and Region &    56.69 & 6.681 $\times 10^{-11}$ &   1.636 $\times 10^{-13}$ \\
World Population by Country and Region without Antarctica &    7.176 & 2.022 $\times 10^{-10}$ &   8.638 $\times 10^{-14}$ \\
\bottomrule
\end{tabular}

    }
\end{table}

\begin{sidewaystable}[htbp]
    \centering
    \caption{Topological Integrity Details}
    \label{tab:intersections_appen}
    \adjustbox{max width=\textwidth}{
    
\begin{tabular}{lrrr|rrr}
\toprule
\multirow{2}{*}{Map} & \multicolumn{3}{c}{Self-intersections} & \multicolumn{3}{c}{Overlap Intersections} \\
\cmidrule(lr){2-4}\cmidrule(lr){5-7}
                   & F4Carto & 5FCarto & BFB & F4Carto & 5FCarto & BFB \\
\midrule

                      Algeria Population by Wilaya (2022) &                           0 &                           0 &                              2 &                            210 &                              0 &                                 0 \\
       Australia Population by State and Territory (2021) &                           0 &                           0 &                              0 &                              8 &                              0 &                                 0 \\
                       Austria Population by State (2020) &                           0 &                           0 &                              1 &                             46 &                              0 &                                 1 \\
                    Bahamas Population by District (2010) &                           0 &                           0 &                             48 &                              7 &                              0 &                                 0 \\
                      Belgium Population by Region (2022) &                           0 &                           0 &                              0 &                              8 &                              0 &                                 0 \\
                        Brazil Population by State (2021) &                           0 &                           0 &                              0 &                             79 &                              0 &                                 0 \\
                      China Population by Province (2020) &                           0 &                           0 &                              0 &                            260 &                              0 &                                 0 \\
              Conterminous USA Population by State (2020) &                           0 &                           0 &                              0 &                             96 &                              0 &                                 0 \\
                     Croatia Covid Cases by County (2022) &                           0 &                           0 &                              0 &                             42 &                              0 &                                 0 \\
                      Croatia Population by County (2021) &                           0 &                           0 &                              0 &                            110 &                              0 &                                 0 \\
                   Egypt Population by Governorate (2017) &                           0 &                           0 &                             57 &                            621 &                              0 &                                54 \\
                      England Population by Region (2022) &                           0 &                           0 &                              3 &                             15 &                              0 &                                 4 \\
              European Union Population by Country (2021) &                           0 &                           0 &                              0 &                            123 &                              0 &                                 0 \\
               European Union Population by Region (2020) &                           0 &                           0 &                             83 &                            375 &                              0 &                               107 \\
          France Population by Metropolitan Region (2008) &                           0 &                           0 &                              0 &                             35 &                              0 &                                 0 \\
                       Germany Population by State (2011) &                           0 &                           0 &                              0 &                            135 &                              0 &                                 0 \\
                  Indonesia Population by Province (2024) &                           0 &                           0 &                              0 &                             41 &                              0 &                                 4 \\
                    Japan Population by Prefecture (2020) &                           0 &                           0 &                              0 &                             24 &                              0 &                                 0 \\
                      Malaysia Population by State (2020) &                           0 &                           0 &                              2 &                             35 &                              0 &                                 0 \\
     Metropolitan France Population by Département (2022) &                           0 &                           0 &                             12 &                            138 &                              0 &                                10 \\
              Russia Population by Federal Subject (2010) &                           0 &                           0 &                             88 &                            389 &                              0 &                                43 \\
           Simplified Belgium Population by Region (2022) &                           0 &                           0 &                              0 &                             13 &                              0 &                                 2 \\
             Singapore Population by Planning Area (2015) &                           0 &                           0 &                            316 &                            483 &                              0 &                               423 \\
                         Switzerland GDP by Canton (2019) &                           0 &                           0 &                              2 &                             91 &                              0 &                                 0 \\
                  Switzerland Population by Canton (2016) &                           0 &                           0 &                              0 &                            137 &                              0 &                                 0 \\
                 Tunisia Population by Governorate (2014) &                           0 &                           0 &                              1 &                             37 &                              0 &                                 0 \\
                          USA Population by County (2021) &                           0 &                           0 &                            195 &                           3400 &                              0 &                               323 \\
                           USA Population by State (2020) &                           0 &                           0 &                             13 &                            227 &                              0 &                                 4 \\
                    Vietnam Population by Province (2019) &                           0 &                           0 &                              0 &                             27 &                              0 &                                 0 \\
                       World Population by Country (2010) &                           0 &                           0 &                             92 &                            117 &                              0 &                                72 \\
                   World Population by Country and Region &                           0 &                           0 &                            891 &                             96 &                              0 &                               474 \\
World Population by Country and Region without Antarctica &                           0 &                           0 &                            237 &                             92 &                              0 &                                55 \\
\bottomrule
\end{tabular}

    }
\end{sidewaystable}

\begin{sidewaystable}[htbp]
    \centering
    \caption{Deformation Details}
    \label{tab:similarity_appen}
    \adjustbox{max width=\textwidth}{
    
\begin{tabular}{lccc|ccc|ccc}
\toprule
\multirow{2}{*}{Map} & \multicolumn{3}{c}{Fréchet Distance} & \multicolumn{3}{c}{Hausdorff Distance} & \multicolumn{3}{c}{Symmetric Difference} \\
\cmidrule(lr){2-10}
                   & F4Carto & 5FCarto & BFB & F4Carto & 5FCarto & BFB & F4Carto & 5FCarto & BFB \\
\midrule

                      Algeria Population by Wilaya (2022) &                      0.83 &                      0.93 &                         0.94 &                        0.47 &                        0.50 &                           0.50 &                          0.92 &                          0.97 &                             0.97 \\
       Australia Population by State and Territory (2021) &                      1.11 &                      0.52 &                         0.54 &                        0.28 &                        0.26 &                           0.26 &                          0.67 &                          0.56 &                             0.57 \\
                       Austria Population by State (2020) &                      0.45 &                      0.30 &                         0.30 &                        0.25 &                        0.22 &                           0.22 &                          0.74 &                          0.50 &                             0.50 \\
                    Bahamas Population by District (2010) &                      1.29 &                      1.23 &                         1.23 &                        0.52 &                        0.58 &                           0.57 &                          0.48 &                          1.12 &                             1.13 \\
                      Belgium Population by Region (2022) &                      0.32 &                      0.33 &                         0.33 &                        0.28 &                        0.21 &                           0.21 &                          0.58 &                          0.35 &                             0.35 \\
                        Brazil Population by State (2021) &                      1.08 &                      0.60 &                         0.60 &                        0.35 &                        0.37 &                           0.37 &                          0.66 &                          0.71 &                             0.71 \\
                      China Population by Province (2020) &                      1.15 &                      0.55 &                         0.55 &                        0.31 &                        0.31 &                           0.31 &                          0.65 &                          0.68 &                             0.68 \\
              Conterminous USA Population by State (2020) &                      0.53 &                      0.48 &                         0.48 &                        0.30 &                        0.31 &                           0.30 &                          0.58 &                          0.58 &                             0.58 \\
                     Croatia Covid Cases by County (2022) &                      0.49 &                      0.46 &                         0.47 &                        0.29 &                        0.27 &                           0.27 &                          0.62 &                          0.59 &                             0.59 \\
                      Croatia Population by County (2021) &                      0.39 &                      0.39 &                         0.39 &                        0.25 &                        0.25 &                           0.25 &                          0.57 &                          0.52 &                             0.53 \\
                   Egypt Population by Governorate (2017) &                      1.25 &                      1.48 &                         1.42 &                        0.48 &                        0.49 &                           0.59 &                          0.87 &                          0.90 &                             0.93 \\
                      England Population by Region (2022) &                      1.10 &                      0.25 &                         0.25 &                        0.32 &                        0.19 &                           0.19 &                          0.65 &                          0.36 &                             0.36 \\
              European Union Population by Country (2021) &                      1.03 &                      0.33 &                         0.33 &                        0.26 &                        0.23 &                           0.23 &                          0.74 &                          0.50 &                             0.50 \\
               European Union Population by Region (2020) &                      1.09 &                      0.72 &                         0.75 &                        0.34 &                        0.43 &                           0.39 &                          0.65 &                          0.78 &                             0.80 \\
          France Population by Metropolitan Region (2008) &                      0.37 &                      0.31 &                         0.31 &                        0.27 &                        0.22 &                           0.22 &                          0.49 &                          0.41 &                             0.41 \\
                       Germany Population by State (2011) &                      0.31 &                      0.25 &                         0.25 &                        0.23 &                        0.22 &                           0.22 &                          0.50 &                          0.47 &                             0.47 \\
                  Indonesia Population by Province (2024) &                      0.68 &                      0.69 &                         0.69 &                        0.39 &                        0.38 &                           0.40 &                          0.85 &                          0.79 &                             0.80 \\
                    Japan Population by Prefecture (2020) &                      1.10 &                      0.59 &                         0.59 &                        0.32 &                        0.33 &                           0.32 &                          0.63 &                          0.73 &                             0.73 \\
                      Malaysia Population by State (2020) &                      0.50 &                      0.55 &                         0.55 &                        0.32 &                        0.35 &                           0.35 &                          0.61 &                          0.63 &                             0.63 \\
     Metropolitan France Population by Département (2022) &                      0.47 &                      0.51 &                         0.51 &                        0.28 &                        0.31 &                           0.30 &                          0.57 &                          0.60 &                             0.60 \\
              Russia Population by Federal Subject (2010) &                      1.22 &                      0.96 &                         0.97 &                        0.37 &                        0.44 &                           0.43 &                          0.60 &                          0.93 &                             0.92 \\
           Simplified Belgium Population by Region (2022) &                      0.57 &                      0.45 &                         0.25 &                        0.28 &                        0.21 &                           0.20 &                          0.59 &                          0.35 &                             0.30 \\
             Singapore Population by Planning Area (2015) &                      1.66 &                      3.58 &                         9.31 &                        0.37 &                        0.70 &                           5.33 &                          0.89 &                          0.79 &                             0.86 \\
                         Switzerland GDP by Canton (2019) &                      0.75 &                      0.68 &                         0.68 &                        0.38 &                        0.37 &                           0.37 &                          0.82 &                          0.78 &                             0.78 \\
                  Switzerland Population by Canton (2016) &                      0.56 &                      0.51 &                         0.51 &                        0.34 &                        0.33 &                           0.33 &                          0.75 &                          0.69 &                             0.69 \\
                 Tunisia Population by Governorate (2014) &                      0.49 &                      0.58 &                         0.58 &                        0.31 &                        0.37 &                           0.37 &                          0.63 &                          0.73 &                             0.73 \\
                          USA Population by County (2021) &                      1.26 &                      1.07 &                         1.50 &                        0.39 &                        0.53 &                           0.67 &                          0.59 &                          0.86 &                             1.00 \\
                           USA Population by State (2020) &                      1.07 &                      0.62 &                         0.62 &                        0.31 &                        0.33 &                           0.33 &                          0.69 &                          0.64 &                             0.64 \\
                    Vietnam Population by Province (2019) &                      1.02 &                      0.44 &                         0.44 &                        0.26 &                        0.29 &                           0.28 &                          0.71 &                          0.61 &                             0.61 \\
                       World Population by Country (2010) &                      1.21 &                      0.89 &                         0.93 &                        0.38 &                        0.48 &                           0.44 &                          0.62 &                          0.81 &                             0.80 \\
                   World Population by Country and Region &                      1.25 &                      1.62 &                         1.67 &                        0.40 &                        0.52 &                           0.49 &                          0.63 &                          0.84 &                             0.84 \\
World Population by Country and Region without Antarctica &                      1.29 &                      0.89 &                         1.00 &                        0.40 &                        0.46 &                           0.47 &                          0.62 &                          0.78 &                             0.81 \\
\bottomrule
\end{tabular}

    }
\end{sidewaystable}

\begin{sidewaystable}[htbp]
    \centering
    \caption{Running-Time Details}
    \label{tab:running_time_appen}
    \adjustbox{max width=\textwidth}{
    
\begin{tabular}{lccc|ccc}
\toprule
\multirow{2}{*}{Map} & \multicolumn{3}{c}{Time (s)} & \multicolumn{3}{c}{\shortstack{Maximum Relative Area \\ Error at Termination}} \\
\cmidrule(lr){2-4}\cmidrule(lr){5-7}
                     & F4Carto & 5FCarto & BFB & F4Carto & 5FCarto & BFB \\
\midrule

                      Algeria Population by Wilaya (2022) &         1.630 &         0.499 &            0.381 &                                    0.302 &                                    0.107 &                                       0.121 \\
       Australia Population by State and Territory (2021) &         0.202 &         0.117 &            0.075 &                                    0.845 &                                    0.718 &                                       0.722 \\
                       Austria Population by State (2020) &         0.249 &         0.084 &            0.064 &                                    0.438 &                                    0.200 &                                       0.199 \\
                    Bahamas Population by District (2010) &         1.767 &         0.367 &            0.422 &                                    6.270 &                                    4.740 &                                       5.330 \\
                      Belgium Population by Region (2022) &         0.117 &         0.043 &            0.037 &                                    0.379 &                                    0.357 &                                       0.357 \\
                        Brazil Population by State (2021) &         0.687 &         0.204 &            0.164 &                                    0.314 &                                    0.186 &                                       0.148 \\
                      China Population by Province (2020) &         1.586 &         0.260 &            0.203 &                                    0.976 &                                    0.743 &                                       0.667 \\
              Conterminous USA Population by State (2020) &         0.564 &         0.126 &            0.089 &                                    0.438 &                                    0.373 &                                       0.431 \\
                     Croatia Covid Cases by County (2022) &         0.445 &         0.077 &            0.065 &                                    0.261 &                                    0.127 &                                       0.130 \\
                      Croatia Population by County (2021) &         0.744 &         0.071 &            0.064 &                                    0.293 &                                    0.090 &                                       0.090 \\
                   Egypt Population by Governorate (2017) &         2.142 &         0.349 &            0.219 &                                    0.698 &                                    0.519 &                                       0.387 \\
                      England Population by Region (2022) &         0.381 &         0.067 &            0.059 &                                    0.548 &                                    0.322 &                                       0.322 \\
              European Union Population by Country (2021) &         1.156 &         0.416 &            0.105 &                                    0.417 &                                    0.292 &                                       0.292 \\
               European Union Population by Region (2020) &         2.059 &         0.349 &            0.155 &                                   16.502 &                                   10.375 &                                      15.529 \\
          France Population by Metropolitan Region (2008) &         0.304 &         0.070 &            0.057 &                                    0.478 &                                    0.278 &                                       0.278 \\
                       Germany Population by State (2011) &         0.575 &         0.082 &            0.073 &                                    0.431 &                                    0.319 &                                       0.329 \\
                  Indonesia Population by Province (2024) &         0.511 &         0.082 &            0.065 &                                    0.978 &                                    0.638 &                                       0.618 \\
                    Japan Population by Prefecture (2020) &         0.347 &         0.173 &            0.107 &                                    0.306 &                                    0.187 &                                       0.143 \\
                      Malaysia Population by State (2020) &         0.194 &         0.085 &            0.045 &                                    0.544 &                                    0.500 &                                       0.483 \\
     Metropolitan France Population by Département (2022) &         1.100 &         0.282 &            0.266 &                                    0.260 &                                    0.163 &                                       0.176 \\
              Russia Population by Federal Subject (2010) &         2.544 &         0.385 &            0.298 &                                    3.300 &                                    2.245 &                                       2.204 \\
           Simplified Belgium Population by Region (2022) &         0.059 &         0.050 &            0.062 &                                    0.223 &                                    0.029 &                                       0.091 \\
             Singapore Population by Planning Area (2015) &         2.110 &         0.185 &            0.175 &                                   29.719 &                                   18.838 &                                      21.457 \\
                         Switzerland GDP by Canton (2019) &         0.624 &         0.145 &            0.084 &                                    0.339 &                                    0.324 &                                       0.327 \\
                  Switzerland Population by Canton (2016) &         0.756 &         0.122 &            0.073 &                                    0.403 &                                    0.268 &                                       0.271 \\
                 Tunisia Population by Governorate (2014) &         0.327 &         0.097 &            0.052 &                                    0.384 &                                    0.342 &                                       0.341 \\
                          USA Population by County (2021) &        10.840 &         2.161 &          239.969 &                                    6.916 &                                    6.646 &                                      12.650 \\
                           USA Population by State (2020) &         1.357 &         0.173 &            0.119 &                                    1.371 &                                    0.778 &                                       0.801 \\
                    Vietnam Population by Province (2019) &         0.338 &         0.131 &            0.082 &                                    0.279 &                                    0.244 &                                       0.237 \\
                       World Population by Country (2010) &         3.024 &         0.448 &            0.366 &                                   32.776 &                                   22.219 &                                      23.595 \\
                   World Population by Country and Region &         4.507 &         0.552 &            0.403 &                                  229.920 &                                  221.658 &                                     219.945 \\
World Population by Country and Region without Antarctica &         3.703 &         0.317 &            0.249 &                                   27.287 &                                   21.302 &                                      22.320 \\
\bottomrule
\end{tabular}

    }
\end{sidewaystable}

\begin{sidewaystable}[htbp]
    \centering
    \caption{Quadtree Target Number of Leaf Nodes \& Running-Time Details}
    \label{tab:quadtree_depth_appen}
    \adjustbox{max width=\textwidth}{
    \begin{tabular}{lrrrrrrrr}
\toprule
Map & \multicolumn{8}{c}{\shortstack{Running time (s) of Cartogram Generation\\ if Quadtree Target Number of Leaf Nodes Chosen\\ as Area of Canvas $\times 2^{-x}$}} \\
\cmidrule(lr){2-9}
     & 5 & 6 & 7 & 8 & 9 & 10 & 11 & 12 \\
\midrule

                      Algeria Population by Wilaya (2022) &   2.0762 &  1.4846 &  1.0330 &  0.9469 &  1.0632 &  1.2804 &  1.6954 &   5.5090 \\
       Australia Population by State and Territory (2021) &   1.6742 &  1.1212 &  0.8130 &  0.7538 &  0.7054 &  0.8801 &  1.1376 &   2.8765 \\
                       Austria Population by State (2020) &   0.7622 &  0.4323 &  0.2696 &  0.2442 &  0.2618 &  0.2653 &  0.3516 &   0.7101 \\
                    Bahamas Population by District (2010) &   4.6396 &  3.5808 &  3.3218 &  3.3607 &  3.8191 &  4.4284 &  5.2772 &  17.0130 \\
                      Belgium Population by Region (2022) &   0.6108 &  0.2862 &  0.1571 &  0.0976 &  0.0772 &  0.0633 &  0.0564 &   0.0942 \\
                        Brazil Population by State (2021) &   1.5295 &  0.9222 &  0.7278 &  0.6374 &  0.7295 &  0.7843 &  0.9357 &   1.4131 \\
                      China Population by Province (2020) &   1.8837 &  1.3087 &  0.9038 &  0.7830 &  0.8824 &  1.0485 &  1.4761 &   2.1226 \\
              Conterminous USA Population by State (2020) &   0.9765 &  0.6451 &  0.4888 &  0.5126 &  0.6321 &  0.8376 &  1.3205 &   2.0654 \\
                     Croatia Covid Cases by County (2022) &   0.7253 &  0.3746 &  0.2879 &  0.2841 &  0.3374 &  0.3574 &  0.5982 &   0.9425 \\
                      Croatia Population by County (2021) &   0.6835 &  0.3678 &  0.2312 &  0.1887 &  0.2478 &  0.3300 &  0.4161 &   0.8249 \\
                   Egypt Population by Governorate (2017) &   2.2748 &  1.3141 &  1.0580 &  0.8976 &  1.0684 &  1.4158 &  2.9497 &   6.6810 \\
                      England Population by Region (2022) &   1.6671 &  1.4498 &  1.3402 &  1.2887 &  1.3265 &  1.3528 &  1.3853 &   1.5047 \\
              European Union Population by Country (2021) &   2.9174 &  2.3958 &  1.9156 &  1.7224 &  1.6306 &  2.7692 &  2.5413 &   6.6767 \\
               European Union Population by Region (2020) &  19.8411 & 12.6082 &  9.1774 &  7.6042 &  4.1199 &  4.9561 &  7.4704 &  11.5696 \\
          France Population by Metropolitan Region (2008) &   0.8217 &  0.4870 &  0.3240 &  0.2674 &  0.2705 &  0.2857 &  0.3031 &   0.5252 \\
                       Germany Population by State (2011) &   0.7414 &  0.4067 &  0.2570 &  0.1755 &  0.2566 &  0.2355 &  0.3647 &   0.7074 \\
                  Indonesia Population by Province (2024) &   0.6961 &  0.4942 &  0.4244 &  0.4186 &  0.5226 &  0.6375 &  1.9409 &   2.1745 \\
                    Japan Population by Prefecture (2020) &   1.3204 &  0.9488 &  0.7157 &  0.7072 &  0.7960 &  1.2052 &  1.1733 &   1.8359 \\
                      Malaysia Population by State (2020) &   0.6019 &  0.3872 &  0.3456 &  0.3370 &  0.4426 &  0.5819 &  0.6604 &   1.0874 \\
     Metropolitan France Population by Département (2022) &   1.4617 &  1.1165 &  0.9957 &  1.1046 &  1.3680 &  1.9062 &  2.2952 &   4.8856 \\
              Russia Population by Federal Subject (2010) &   5.0835 &  2.4048 &  1.6316 &  1.5357 &  2.0063 &  3.6215 &  4.5327 &   9.8079 \\
           Simplified Belgium Population by Region (2022) &   0.5824 &  0.2706 &  0.1296 &  0.0760 &  0.0553 &  0.0432 &  0.0481 &   0.0613 \\
             Singapore Population by Planning Area (2015) &   6.9973 &  8.2264 &  4.1267 &  2.3801 &  3.0192 &  3.4972 &  5.7284 &  11.5871 \\
                         Switzerland GDP by Canton (2019) &   1.4040 &  0.6426 &  0.4508 &  0.4060 &  0.4366 &  0.6230 &  0.8561 &   1.3759 \\
                  Switzerland Population by Canton (2016) &   0.8469 &  0.5254 &  0.3762 &  0.3430 &  0.3792 &  0.4506 &  0.7366 &   0.9596 \\
                 Tunisia Population by Governorate (2014) &   0.7579 &  0.5007 &  0.3480 &  0.3544 &  0.4438 &  0.5593 &  0.7215 &   1.6400 \\
                          USA Population by County (2021) & 142.1828 & 61.4285 & 44.8844 & 43.0581 & 44.6720 & 56.7868 & 91.0163 & 110.2354 \\
                           USA Population by State (2020) &   1.3565 &  0.9503 &  0.7965 &  0.7859 &  0.8823 &  1.3025 &  2.2353 &   4.1051 \\
                    Vietnam Population by Province (2019) &   0.7406 &  0.5877 &  0.5957 &  0.6948 &  0.9948 &  1.3048 &  1.6782 &   2.4427 \\
                       World Population by Country (2010) &  13.4673 & 11.0617 &  6.0427 &  5.6023 &  5.6853 &  8.1085 & 12.0031 &  17.6184 \\
                   World Population by Country and Region &  20.3101 & 10.0227 & 10.9023 &  8.9238 & 11.0616 & 18.7876 & 13.8273 &  15.2280 \\
World Population by Country and Region without Antarctica &  14.6730 & 10.5958 &  7.4114 &  5.8852 &  7.5664 & 11.9035 & 15.8954 &  23.0149 \\
\bottomrule
\end{tabular}

    }
\end{sidewaystable}

\begin{figure}[htbp]
    \centering
    \includegraphics[width=\textwidth, keepaspectratio]{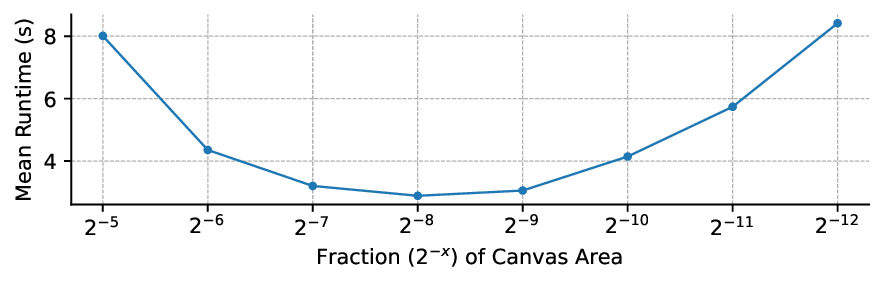}
    \caption{Impact of the target quadtree leaf cell count on cartogram running time.}
    \label{fig:quadtree_depth}
\end{figure}

\end{document}